\documentclass[a4paper,fleqn,usenatbib]{mnras}

\usepackage{url}
\usepackage[textwidth=1.5cm,shadow]{todonotes}	% \todo{}: side notes at the margins; \todo[inline]{}: inline notes; \missingfigure{}; \listoftodos
\usepackage[normalem]{ulem}    
\usepackage{color,soul}

\usepackage{graphicx}
\usepackage{rotating}
\usepackage{verbatim}
\usepackage{wrapfig}	% wrap text around figures; https://en.wikibooks.org/wiki/LaTeX/Floats,_Figures_and_Captions
\usepackage{mdframed}	% put simple frame around part of the text; http://tex.stackexchange.com/questions/36524/how-to-put-a-framed-box-around-text-math-enviroment

\newcommand{\fermi}{\emph{Fermi}}

\newcommand{\orcid}[1]{\href{#1}{\includegraphics[scale=0.035]{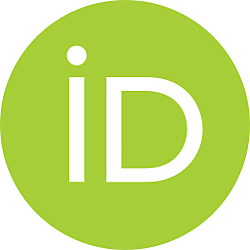}}}

\definecolor{boxback}{HTML}{DAE5F0}
\definecolor{boxframe}{HTML}{0F365B}
{
\end{tcolorbox}   
}

\newcommand{\eg}[1]{(e.g. \citealt{#1})}

\title[Jet efficiencies and spins of FSRQs]{Jet efficiencies and black hole spins in jetted quasars}

\author[Soares \& Nemmen]{
Gustavo Soares$^1$\thanks{E-mail: soaresgrr@gmail.com}\orcid{https://orcid.org/0000-0001-5489-4925}
and
Rodrigo Nemmen$^1$\orcid{https://orcid.org/0000-0003-3956-0331}
\\
$^1$Universidade de S\~ao Paulo, Instituto de Astronomia, Geof\'{\i}sica e Ci\^encias Atmosf\'ericas, Departamento de Astronomia,\\ S\~ao Paulo, SP 05508-090, Brazil\\}

\date{Accepted XXX. Received YYY; in original form ZZZ}

\pubyear{2019}

\begin{document}
\label{firstpage}
\pagerange{\pageref{firstpage}--\pageref{lastpage}}
\maketitle

\begin{abstract}
The mechanisms responsible for the production of relativistic jets from supermassive black holes (SMBHs) accreting at near-Eddington rates are not well-understood. Simple theoretical expectations indicate that SMBHs in quasars accrete via thin discs which should produce at most very weak jets. This is contradicted by observations of powerful jets in flat-spectrum radio quasars (FSRQs). We use gamma-ray luminosities observed with the \fermi\ Large Area Telescope as a proxy of the jet power for a population of 154 FSRQs. Assuming typical quasar accretion rates and using black hole mass measurements from a variety of methods, we find a mean jet production efficiency of about 10 per cent for FSRQs, with values as high as 222 per cent. We find that this is consistent with FSRQs hosting moderately thin, magnetically arrested accretion discs around rapidly spinning black holes (BHs). Modeling our observations using general relativistic magnetohydrodynamic (GRMHD) simulations of jets from thin discs, we find an average lower limit of $a_* = 0.59$ for the SMBH spins of FSRQs, with tendency for the spins to decrease as the black hole mass increases. Our results are consistent with the merger-driven evolution of SMBHs. 3 per cent of the sample cannot be explained by current GRMHD models of jet production from Kerr BHs due to the high efficiencies. Along the way, we find a correlation between BH masses and $L_\gamma$ which may be an useful mass estimator in blazar gamma-ray studies.
\end{abstract}

\begin{keywords}
galaxies: jets -- gamma-rays: general -- black hole physics -- accretion discs -- galaxies: active
\end{keywords}

\section{Introduction}{\label{sec:intro}}

Blazars are among the most powerful active galactic nuclei (AGN), producing relativistic jets roughly aligned with the line of sight of the observer. As a result, they become strongly Doppler boosted, thereby allowing the jet emission to dominate the spectrum of blazars. Based on their observational properties, it is customary to divide blazars into two different categories: flat-spectrum radio quasars (FSRQs) are characterised by significant broad emission lines observed in their spectra, while blazars with barely any noticeable emission lines are called BL Lacertae (BL Lac) objects \citep{Urry95,Massaro2016}.

The extended radio structure and large radio luminosities of FSRQs suggest that Fanaroff-Riley Type II (FR II, \citealt{Fanaroff74}) radio galaxies are the parent populations of radio-loud quasars, whereas FR I galaxies are the parent populations of BL Lacs \citep{Urry95}. In view of this, \cite{Bottcher02} and \cite{Cavaliere02} proposed evolutionary models of blazars, concluding that FSRQs would evolve into BL Lacs as their accretion decreases over time, which is supported by the findings in \cite{Ajello13}. 

FSRQs are conventionally interpreted as supermassive black holes (SMBHs) accreting at high mass accretion rates $\dot{M} \ga 0.01\; \dot{M}_{\rm Edd}$ and are therefore surrounded by thin accretion discs ($\dot{M}_{\rm Edd}$ is the Eddington accretion rate); BL Lacs have $\dot{M} \la 0.01\;\dot{M}_{\rm Edd}$, hence their SMBHs are fed via radiatively inefficient accretion flows (RIAFs) \eg{Ghisellini2009a, Ghisellini2014}. 

According to standard models, the formation of jets of relativistic jets is a function of mainly two parameters: the mass accretion rate in units of the Eddington rate $\dot{m} \equiv \dot{M} / \dot{M}_{\rm Edd}$ and the dimensionless black hole (BH) spin $a_* \equiv a/M$ \eg{Contopoulos2014}. For instance, the Blandford-Znajek (BZ) mechanism  \citep{Blandford77}) implies that it is possible to extract rotational energy of Kerr BHs threaded by large-scale magnetic field lines which are brought to their vicinity by the accreted gas. Therefore, the magnetic field -- in practice represented by the magnetic flux $\phi$ at the event horizon -- is another fundamental parameter to be considered when studying jet formation \eg{Semenov2004}.

FSRQs and FR II radio galaxies produce bright, powerful jets and should have standard, geometrically thin, optically thick accretion discs \eg{Ghisellini2001, Jester2005, Ghisellini2009a}. Therefore, observationally it is clear that stronger jets associated with FSRQs can be produced by thin discs, contrary to some theoretical expectations \eg{Meier2001,Nemmen2007}. General relativistic magnetohydrodynamic (GRMHD) simulations are now part of the standard machinery of tools used to study the dynamics and electromagnetic appearance of accreting BHs \eg{Porth2019}. GRMHD models have been advancing tremendously due to a combination of algorithmic advances and Moore's law. The state of the art in terms of jet formation models in high-$\dot{M}$ BH systems involve the addition of radiation pressure effects in the numerical models: general relativistic radiative MHD (GRRMHD; \citealt{Sadowski2014,McKinney2014, Ryan2015}). Such models are beginning to reach a level where they can make testable predictions. Therefore, it is worth assessing the state of the numerical models when confronted with observations of jetted quasars, and what constraints on the model parameters can be obtained from such comparisons.

GRMHD models have demonstrated that magnetically arrested discs (MAD; \citealt{Sasha11,McKinney12}) maximise jet power. The reason is that in a MAD, $\phi$ accumulates near the event horizon and is maximised, being prevented from escaping due to the pressure exerted by the inflowing gas. The jet power $P$ is given by $P \propto ( a_* \Phi/M )^2$ to first order, hence if $\phi$ is maximised for a given $a_*$, $P$ will also be maximal.

Given that FSRQs produce extremely powerful jets, MADs naturally emerge as an explanation. \cite{Ghisellini2014} found that a strong correlation between jet power and accretion luminosity with the jet power dominating the disc luminosities, in agreement with MAD expectations. \cite{Zamaninasab2014} found in a sample dominated by FSRQs that the estimated values of $\phi$ are consistent with MAD predictions. The MAD hypothesis was questioned by \cite{van-Velzen13}, who studied a sample of FR II quasars and found a jet production efficiency $\eta \approx 0.01$ -- considerably lower than MAD predictions and other observational estimates  \citep{Ghisellini2014,Zamaninasab2014,Zdziarski15}.

Many GR(R)MHD works have investigated thin discs and their energy outflows  \citep{Shafee08b,Noble09,Noble10,Penna10,Noble11,Avara2016,Liska19}. In particular, \cite{Avara2016} performed a systematic numerical investigation of the radiative and jet efficiencies of thin MADs around Kerr BHs as a function of $a_*$. Besides finding that thin MADs achieve a radiative efficiency of 15 per cent which is twice the Novikov-Thorne value, they fitted a function to jet production efficiency resulting from several models and proposed a function $\eta=\eta(a_*, \phi, h)$ which is related to the jet power as $P = \eta(a_*, \Phi) \dot{M} c^2$ and $h$ is the disc thickness $h \equiv H/R$ where $H$ is the scale height. The advantage of this model and the simplicity of the fitting function for $\eta$ is that one can quickly check whether the typical jet efficiencies predicted by GRMHD MAD models can explain observations of quasars, as well as performing quantitative constraints on the BH parameters such as the spin.

Indeed, the Avara et al. model was applied to observations of FR II quasars by \cite{Rusinek17}, who found that the jets in their sample could only be explained by the MAD scenario if the discs are thicker than predicted in standard theory. \cite{Inoue17} analysed a large sample of more than $7000$ radio-loud quasars detected at $1.4\;\mathrm{GHz}$ and SDSS optical spectra and found jet efficiencies comparable to those suggested by \cite{van-Velzen13}, and a low average spin of $a_* = 0.13$, which is smaller than cosmological merger SMBH evolution models \citep{Volonteri07,Volonteri13}.

Previous works such as \cite{Rusinek17,Inoue17} have based their samples on radio and optical observations of radio-loud quasars. FSRQs are bright gamma-ray emitters and are behind the majority of point sources observed by the gamma-ray instruments. For instance, the \fermi\ Large Area Telescope (LAT) has observed more than 600 FSRQs in the 0.1-300 GeV energy range \citep{3fgl,4fgl}. Here, we will use \fermi\ LAT observations of FSRQs which constrain their energetics -- concretely, their jet powers and jet production efficiencies -- with the goal of assessing how well current models for jet formation from high-$\dot{M}$ BHs are able to explain such observations. Along the way, we will obtain quantitative estimates of BH spins for jetted quasars, which may shed light on the spin evolution of SMBHs.

This paper is organised as follows. Section \ref{sec:data} discusses the gamma-ray observations and SMBH masses. Section \ref{sec:res} contains the results, which are discussed in Section \ref{sec:disc}. We present a summary in Section \ref{sec:conc}. For all our cosmological calculations, we use $H_0 = 67.74\;\mathrm{km}\;\mathrm{Mpc}^{-1}\;\mathrm{s}^{-1}$, $\Omega_\mathrm{m_0} = 0.31$, and $\Omega_{\Lambda_0} = 0.69$  \citep{Planck15}.

\section{Data selection}	\label{sec:data}

For our study, we need a sample of jetted quasars with estimated jet powers and BH masses. This sample was provided by \cite{Ghisellini2014, Ghisellini15} (hereafter G14 and G15, respectively). 

G14 and G15 published a sample of blazars that have been detected in $\gamma$-rays by \fermi-LAT and spectroscopically observed in the optical band \citep{Shaw12,Shaw13}, including 229 FSRQs and 475 BL Lacs. The Ghisellini sample is based on the first and second \fermi-LAT catalogs, corresponding to only two years of $\gamma$-ray observations. We have now more than ten years of LAT observations, therefore we cross-matched the original sample of FSRQs from G14 and G15 with the most up-to-date catalog of LAT sources -- 4FGL; \citealt{4fgl} -- ending up with 191 FSRQs.

We now describe our procedure for matching the FSRQs to the 4FGL catalog. Since different catalogs adopt different naming conventions for their objects and report slightly different declinations and right ascensions, matching the FSRQs in G15 with the FSRQs detected by \fermi 4FGL using some kind of direct comparison between the object names, their aliases in either catalog or their position in the sky is not very effective. To overcome this, we employed a ``distance method''. For each object in G15, we compared its coordinates with those of the objects in \fermi 4FGL, then calculated the separation $d$ given by
\begin{equation}
    d = \sqrt{[(\mathrm{\alpha_G} - \mathrm{\alpha_F})\cos(\delta_F)\cos(\delta_G)]^2 + (\mathrm{\delta_G} - \mathrm{\delta_F})^2},
    \label{eq:distance_method}
\end{equation}
between this object and each object in \fermi 4FGL. We then selected the smallest distance, $d_\mathrm{min}$, and compared it to a threshold distance $d_\mathrm{thr}$. If $d_\mathrm{min} \leq d_\mathrm{thr}$, we concluded that the objects match. In Eq. \ref{eq:distance_method}, the subindices $\mathrm{F}$ and $\mathrm{G}$ refer to the \fermi 4FGL and the G15 objects, respectively, the declination $\delta$ and right ascension $\alpha$ are given in radians. After some trial and error, we settled on a threshold value $d_\mathrm{thr}$ corresponding to one minute. Hence, $d \leq 1$ minute implied a match between the objects.

Using this method, we were able to identify 156 objects in the \fermi 4FGL catalog, out of the 191 sources in G15. Still, the objects 1438+3710 and 1439+3712 in G15 were both associated with 4FGL J1438.9+3710 by our distance method. Similarly, the objects 1636+4715 and 1637+4717 were both associated with 4FGL J1637.7+4717. Given their significantly different redshifts, we searched the literature and the 4FGL and G15 aliases to determine which object should be correctly associated with both 4FGL sources, and we determined that both 1439+3712 and 1636+4715 should be excluded from our analysis. Hence, our final sample contains 154 objects. Table \ref{tab:data} lists the FSRQs names, coordinates along with the other relevant properties for this work such as their $\gamma$-ray luminosities and BH masses. 

\begin{figure}
\includegraphics[width=\linewidth]{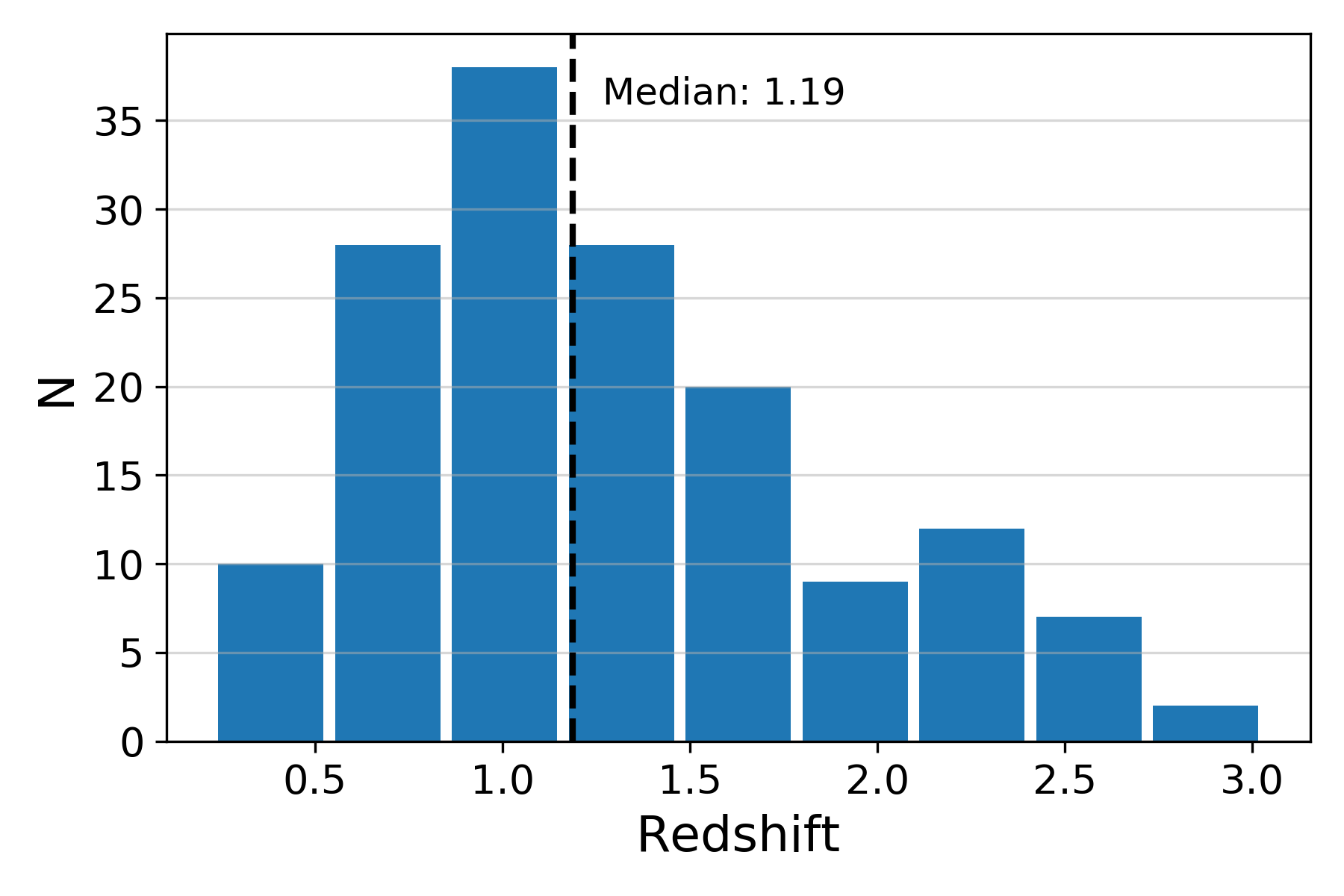}
\caption{Redshift distribution of the 154 FSRQs in our sample. }
\label{fig:z_hist}
\end{figure}

The remaining objects from G15 which remained unmatched with 4FGL objects could possibly be due to sources which were bright enough for a few months to be included in previous \fermi LAT catalogs but not over the entire 8-year period comprised by \fermi 4FGL \eg{Paliya17}. Furthermore, at least one object in G15 appears listed as BL Lac elsewhere. Therefore, we chose to stick to our ``distance method'' based on Eq. \ref{eq:distance_method} to select our sample, and the non-identification of the remaining objects does not compromise the data or the results presented in this work. The redshifts of the FSRQs considered here vary between $0.226$ and $3$, with a mean value of $1.31$ (Figure \ref{fig:z_hist}).

\begin{figure}
\includegraphics[width=\linewidth]{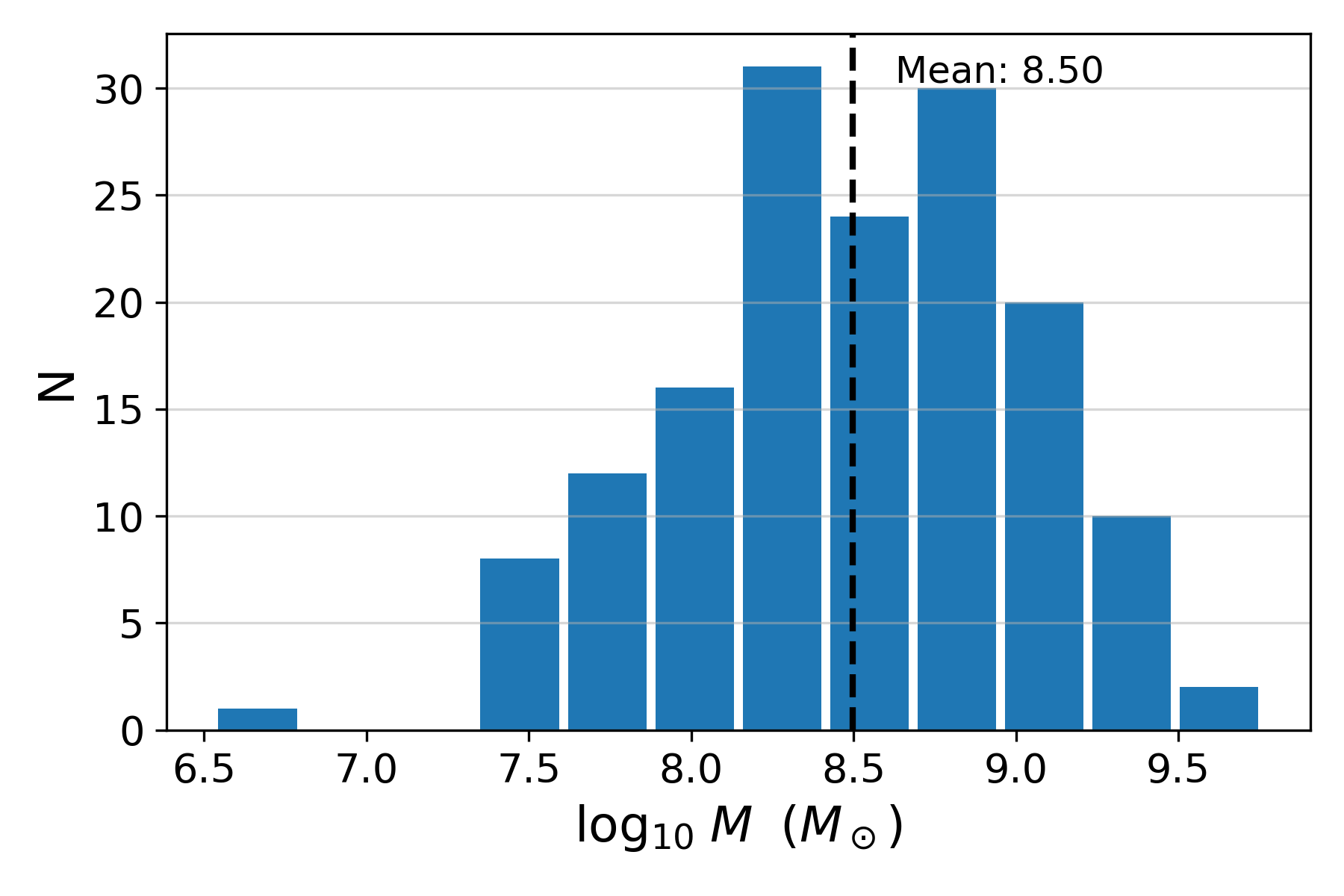}
\caption{Distribution of the black hole masses $M$ in our sample.}
\label{fig:mbh_hist}
\end{figure}

G15 estimated the BH masses using four different methods. Three of these are virial estimates based on emission lines measured by \cite{Shaw12}, whereas the other uses a disc-fitting method based on the \cite{Shakura1973} thin disc model. In this work, we compute the BH masses as the average between the virial estimates, ignoring the disc-fitting estimates altogether because this method could not be applied to all FSRQs in the original sample of G15 due to poor data or overdominance of the synchrotron jet component \citep{Ghisellini15}. The uncertainty of the mass measurements obtained by the virial method is $0.5\;\mathrm{dex}$. Fig. \ref{fig:mbh_hist} displays the mass distribution of the objects in our sample.

\begin{figure}
\includegraphics[width=\linewidth]{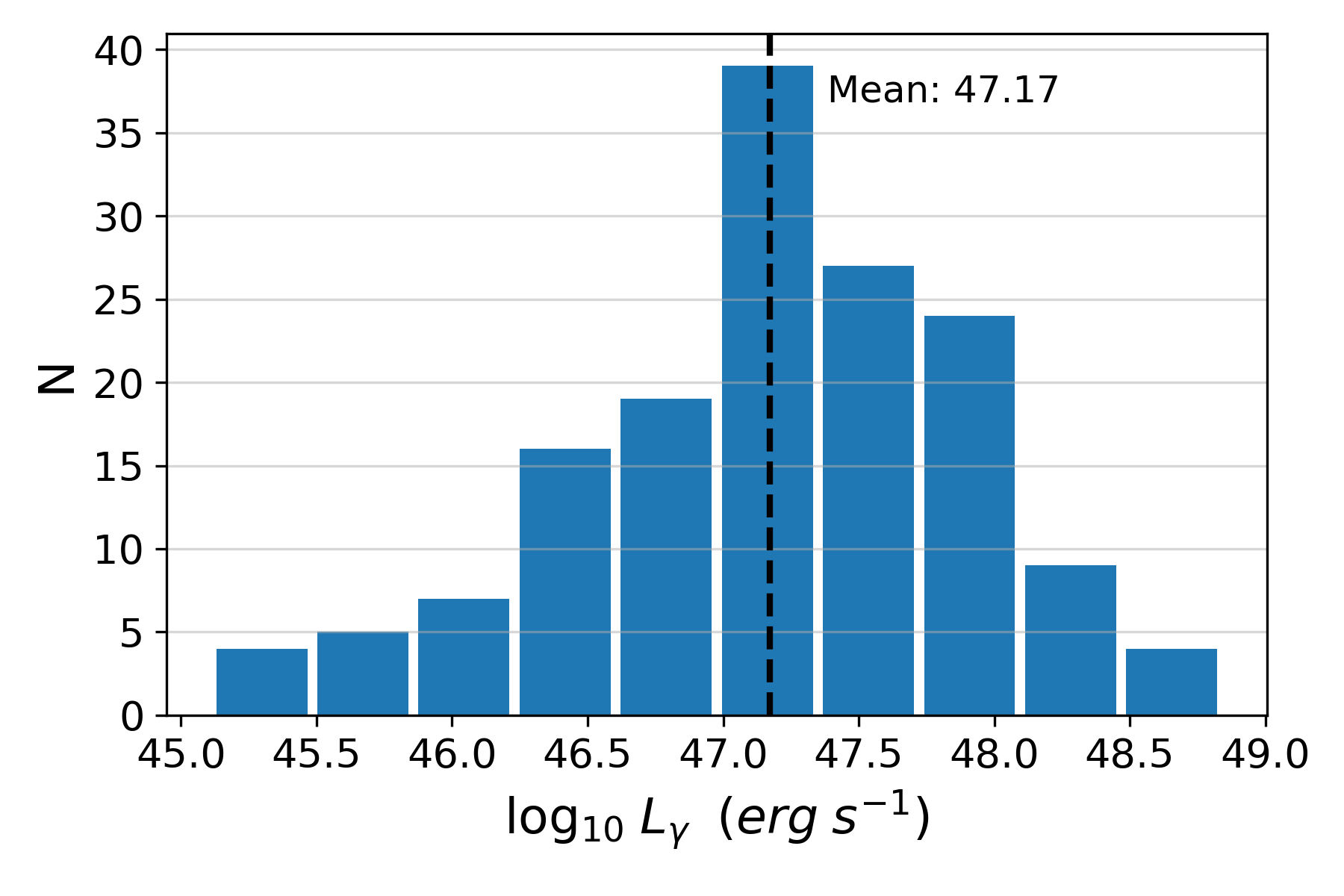}
\caption{Distribution of the gamma-ray luminosities in our sample.}
\label{fig:lg_hist}
\end{figure}

The gamma-ray luminosities $L_\gamma$ observed with \fermi-LAT are shown in  Fig. \ref{fig:lg_hist}. They were estimated from the 4FGL energy fluxes $F_\gamma$ in the $100\;\mathrm{MeV}$ to $100\;\mathrm{GeV}$ range obtained assuming a power-law model: 
\begin{equation}
    L_\gamma = 4\pi d_L^2\frac{F_\gamma}{(1+z)^{(1-\alpha_g)}},
    \label{eq:lgamma}
\end{equation}
where $d_L$ is the luminosity distance and $\alpha_g = 1 - n$ is the energy spectral index, with $n$ being the power-law spectral index.

\section{Results}   \label{sec:res}

\begin{figure}
\includegraphics[width=\linewidth]{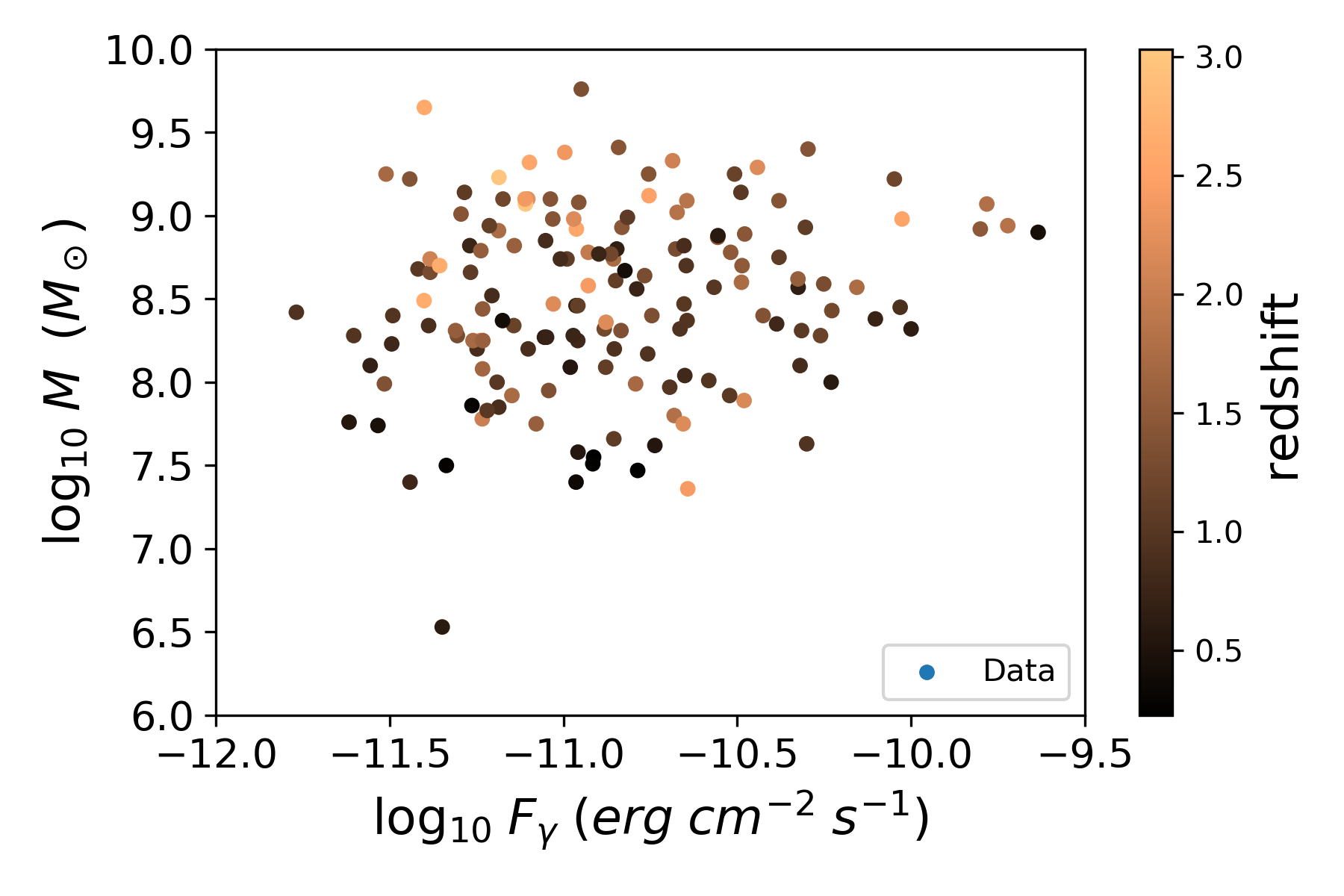}
\caption{Black hole mass estimates collected from G15 plotted against the \fermi-LAT gamma-ray flux.}
\label{fig:mass-flux}
\end{figure}

Figure \ref{fig:mass-flux} shows the gamma-ray flux plotted against the black hole masses for all objects in our sample. We find only a weak level of correlation between $F_\gamma$ and $M$: the Pearson correlation coefficient is $r = 0.15$ resulting in a probability of no-correlation of $p_{\rm null}=0.07$ (two-tailed \textit{p}-value) -- a high chance of the result being consistent with the null hypothesis.

\begin{figure}
\includegraphics[width=\linewidth]{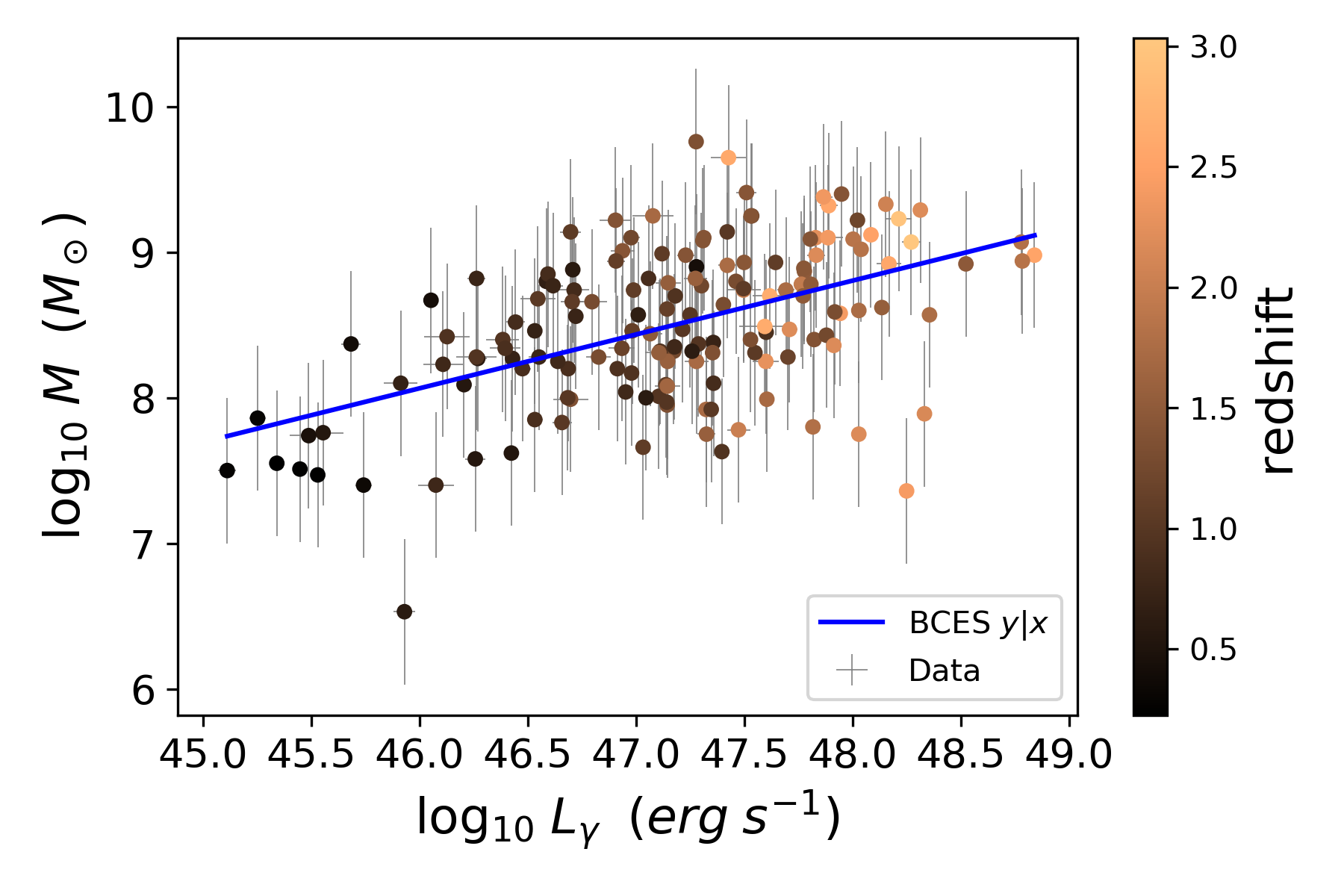}
\caption{Black hole mass plotted against the \fermi-LAT gamma-ray luminosity. The uncertainties in $L_\gamma$ are too small and barely visible, while the uncertainties in the black hole mass are much larger. The solid blue line corresponds to our best-fit line.}
\label{fig:mass_lg}
\end{figure}

Figure \ref{fig:mass_lg} shows the gamma-ray luminosity as a function of the black hole mass and redshift $z$ for all objects in our sample. On average, the FSRQs with higher gamma-ray luminosities tend to be located at higher redshifts \eg{Ajello2012}. Figure \ref{fig:mass_lg} suggests a possible correlation between the BH mass and $L_\gamma$. Indeed, the Pearson correlation coefficient is $r = 0.5$ resulting in a probability of no-correlation of $p_{\rm null}=10^{-11}$. 

Given the above value of $p_{\rm null}$, it is tempting to conclude that there is a strong correlation between these two variables. This would be puzzling given the weak correlation between the flux and $M$ in Fig. \ref{fig:mass-flux}. To settle the issue, we performed a partial correlation analysis of the common dependence of $L\gamma$ and $M$ on $d_L$ using the partial Kendall's tau correlation test \citep{Akritas1996}. Considering the log of these values, we find that the p-value of the null hypothesis (i.e. no correlation between $L\gamma$ and $M$ in the presence of the third variable $d_L$) is very low (Table \ref{tab:partial}). Therefore, the $L_\gamma-M$ correlation is unlikely to be entirely a distance-driven artifact. The fact that $M$ is moderately correlated with $z$ ($r=0.43$, $p_{\rm null}=10^{-8}$), combined with $L_\gamma \propto d_L^{-2}$ explains why the $M-L_\gamma$ correlation is stronger than the $M-$Flux one: Fig. \ref{fig:mass_lg} is showing two variables that depend on the distance to different degrees.

\begin{table*}
\caption{Results of partial correlation analysis. Columns (1)-(2): Quantities (Z is $\log_{10} d_\mathrm{L}$ in both cases). Column (3): Subsample; Column (4): Number of sources; Columns (5)-(7): results of partial correlation analysis; $\tau$ is the partial Kendall’s correlation coefficient; $\sigma$ is the square root of the calculated variance; $P_{\rm null}$ is the probability for accepting the null hypothesis that there is no correlation between X and Y; Column (7) gives the associated significance in standard deviations with which the null hypothesis is rejected.}
\label{tab:partial}
\begin{tabular}{@{} *{7}c @{}}
\hline
X & Y & Objects & N & $\tau$ & $\sigma$ & $P_\mathrm{null}$\\
\hline
$\log_{10} L_\gamma$ & $\log_{10} M$ & all & 150 & 0.21 & 0.04 & $5 \times 10^{-7}$ ($5 \sigma$)\\
\hline
\end{tabular}
\end{table*}

We fitted a linear relation 
\begin{equation}	\label{eq:lg_mass}
\log_{10}M = A\log_{10}L_\gamma + B
\end{equation}
to our data using the BCES $y|x$ method \citep{Akritas1996bces} with $M$ in solar masses and $L_\gamma$ in $erg\;s^{-1}$. The BCES $y|x$ method is a regression procedure that takes into account uncertainties in both variables and assumes that $M$ is the dependent variable -- something justifiable in this case given the much larger uncertainties affecting $M$. The best-fit parameters resulting from the fit are $A = 0.37 \pm 0.05$ and $B = -8.95 \pm 2.42$. The corresponding fit is indicated in Fig. \ref{fig:mass_lg}. The scatter about the best-fit is 0.5 dex. This correlation between BH mass and $L_\gamma$ may be an useful mass estimator in blazar $\gamma$-ray studies in the absence of other, more conventional mass proxies such as broad emission lines.

\cite{Nemmen12} obtained a tight correlation between $L_\gamma$ and the total jet power of blazars:
\begin{equation}
    \log_{10}P_\mathrm{jet} = (0.51 \pm 0.02)\log_{10}L_\gamma + (21.2 \pm 1.1)
    \label{eq:pjet}
\end{equation}
with both variables in units of $erg\;s^{-1}$. Here, we use this correlation to directly estimate the jet power from the measured values of $L_\gamma$. In \cite{Nemmen12}, the scatter around the best-fit is $0.5\;\mathrm{dex}$, which we take as the uncertainty in our $P_\mathrm{jet}$ estimates. The resulting mean of $\log_{10} P_\mathrm{jet}$ is $45.26$.

In order to estimate the efficiency of jet production $\eta = P_\mathrm{jet}/(\dot{M}c^2)$, we need mass accretion rates for the quasars in the our sample. An excellent assumption \eg{McLure2004,Ghisellini2010} is to simply assume that the SMBHs in quasars are accreting at a level of 10 per cent the Eddington rate, i.e. 
take $\dot{M} = 0.1\dot{M}_\mathrm{Edd} = L_\mathrm{Edd} / c^2$ where $L_\mathrm{Edd}$ is the Eddington luminosity. 

\begin{figure}
\includegraphics[width=\linewidth]{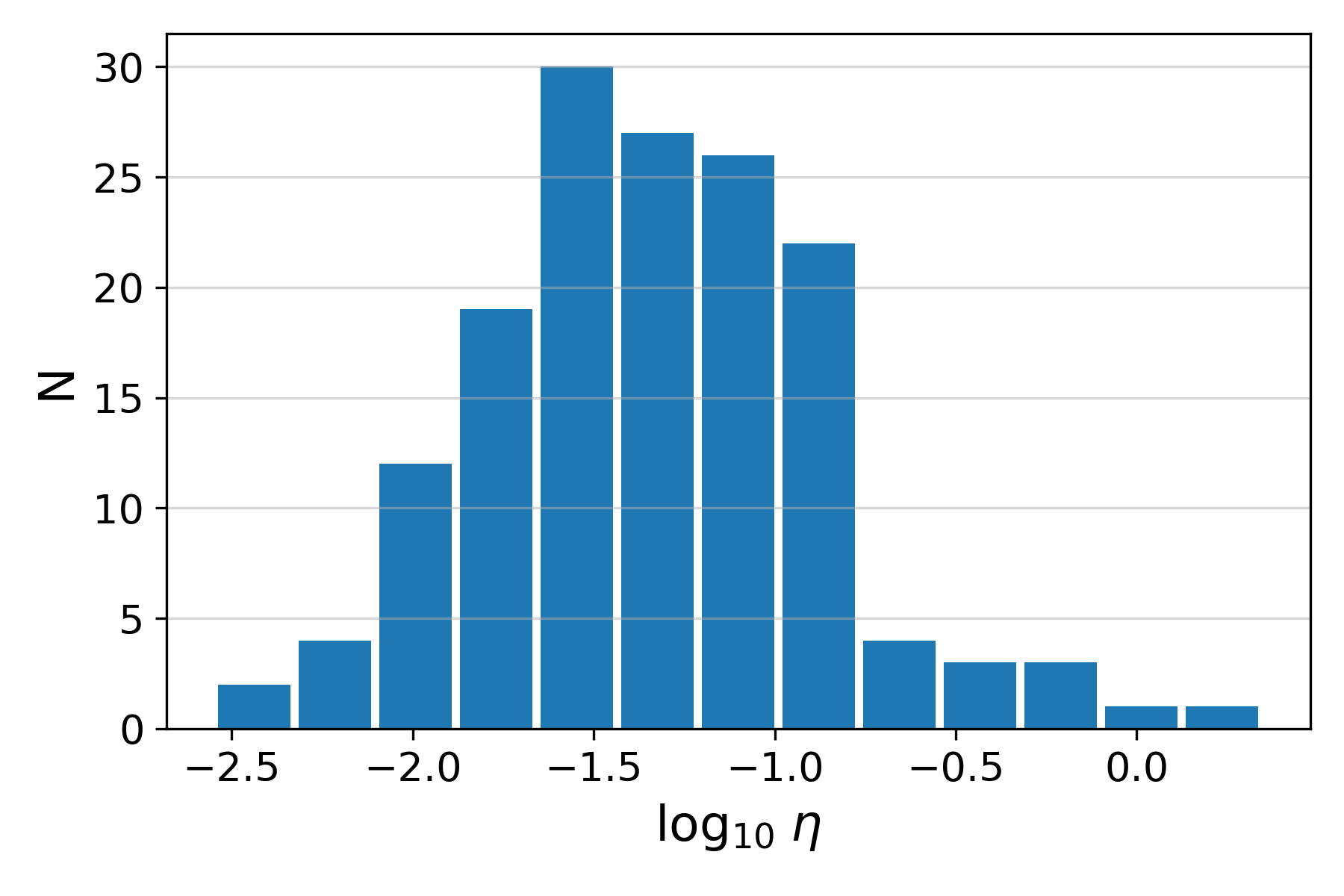}
\caption{Distribution of the jet production efficiencies for the FSRQs in our sample assuming that the SMBHs are accreting at 10 per cent of the Eddington rate.}
\label{fig:eta_hist}
\end{figure}

The distribution of observationally-constrained jet production efficiencies for the objects in our sample is plotted in Fig. \ref{fig:eta_hist}. We note that the $0.5\;\mathrm{dex}$ uncertainties in both $P_\mathrm{jet}$ and $M$ imply large uncertainties in $\eta$ of about $0.7$ dex. The median efficiency in our sample is $0.04$; the minimum and maximum efficiencies are, respectively, about $3 \times 10^{-3}$ and $2$.

\begin{figure}
\centering
\includegraphics[width=\linewidth]{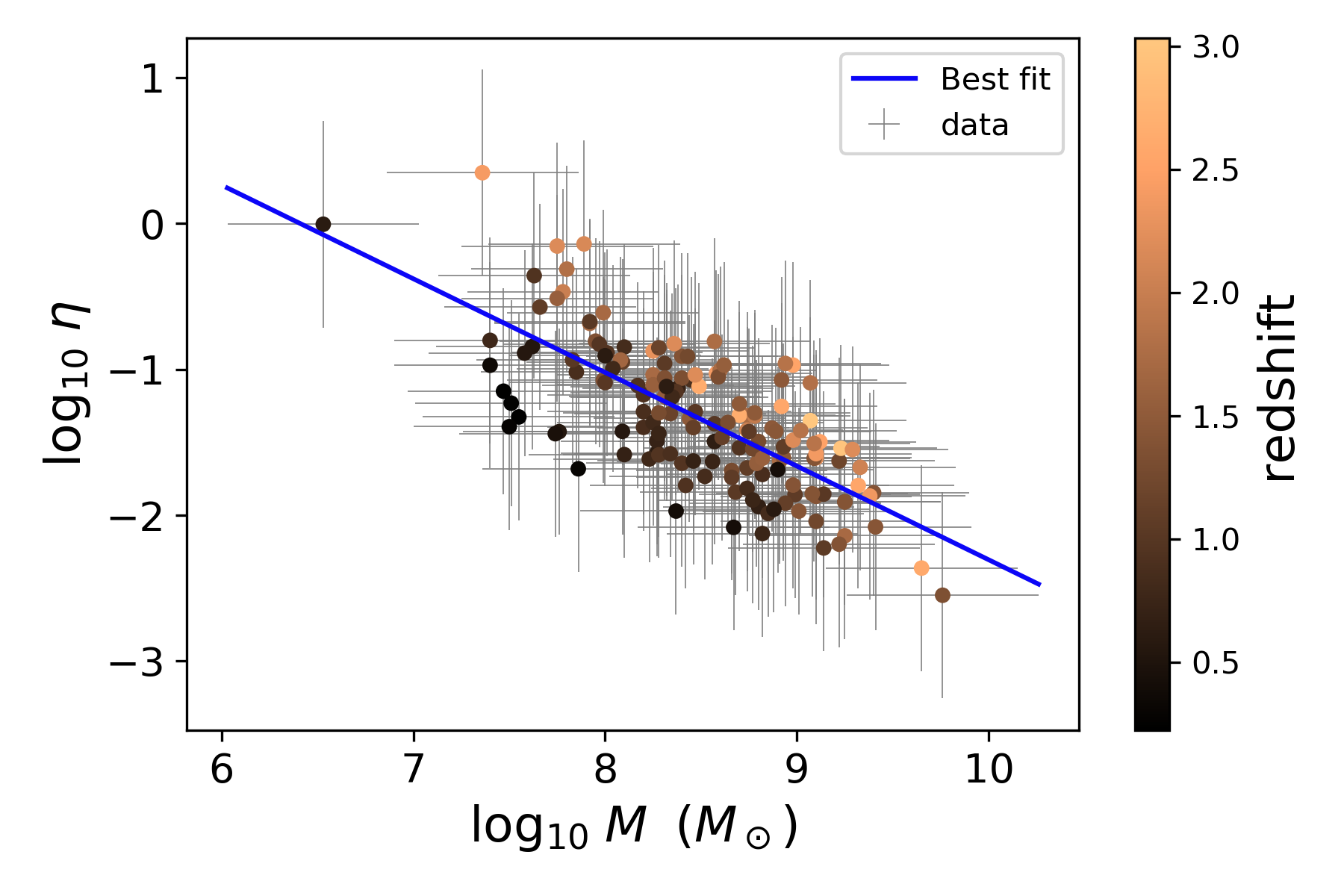}
\caption{Jet production efficiencies as a function of BH mass. See caveats mentioned in section \ref{sec:res}.}
\label{fig:eta-mass}
\end{figure}

Figure \ref{fig:eta-mass} displays the relation between $\eta$ and the BH mass. By construction, $\eta \propto M^{-1}$ due to our assumption of $\dot{M} \propto \dot{M}_{\rm Edd}$. Therefore, one must be careful in interpreting any possible correlations between $\eta$ and $M$. Given this important caveat and keeping in mind the considerable uncertainties affecting our estimates of $\eta$, we find an anticorrelation between $\eta$ and $M$,
\begin{equation} % log eta = -0.65*log M/Msun + 4.17
\eta = 0.1 \left( \frac{M}{10^8 M_\odot} \right)^{-0.64}.
\label{eq:eta-mass}
\end{equation}
In order to make more reliable assessments of the relation between $\eta$ and $M$, we need independent constraints on these variables, i.e. obtain $\eta$ independently from $M$.

\subsection{Black hole spins} \label{sec:res_spins}

We searched the literature for the state-of-the-art models capable of explaining the launching of relativistic jets from thin accretion discs, as appropriate for jetted quasars. The most promising model for FSRQs corresponds to the GRMHD simulations of moderately thin MADs carried out by \cite{Avara2016}. Using data obtained from a variety of disc thicknesses, \cite{Avara2016} obtained an empirical expression for the jet production efficiency,
\begin{equation}
    \eta_{\rm model} \approx 4 \omega_H^2 \left(1 + \frac{0.3\omega_H}{1 + 2 h^4} \right)^2 h^2,
    \label{eq:eta_avara}
\end{equation}
where $\omega_H \equiv a_*/r_H$ is the BH rotation frequency, $r_H = 1 + \sqrt{1 - a_*^2}$ is the horizon radius, $h \equiv \arctan(c_\mathrm{s}/v_\mathrm{\phi}) \approx H/R$ is the disc thickness, $c_\mathrm{s}$ is the sound speed and $v_\mathrm{\phi}$ is the rotational speed \citep{Avara2016,McKinney12}. The dependence $\eta \propto (H/R)^2$ was also reported in \cite{Penna10}.
Thus we have a model that provides a full mapping of the observed jet efficiency to the spin and magnetic flux 
\begin{equation}    \label{eq}
\eta=\eta_{\rm model}(a_*, \phi).
\end{equation}
We proceed by solving this nonlinear equation, using the values of $\eta$ displayed in Figure \ref{fig:eta_hist} to constrain the BH spin--assuming of course that the Kerr metric is the correct description of the spacetime. When solving equation \ref{eq}, we follow \cite{Avara2016} and adopt $h = 0.13$. 

We found that 36 objects -- 23 per cent of the sample -- require $\eta$ larger than the maximum value allowed by eq. \ref{eq:eta_avara}, 
\begin{equation}
\eta > \rm{max} (\eta_{\rm model}) = \eta_{\rm model}(a_*=0.998) = 0.098.
\end{equation}
where the maximum allowed spin is $\rm{max} (a_*) = 0.998$ \citep{Thorne74}. We discuss possible reasons for such large efficiencies in the next section.

\begin{figure}
\includegraphics[width=\linewidth]{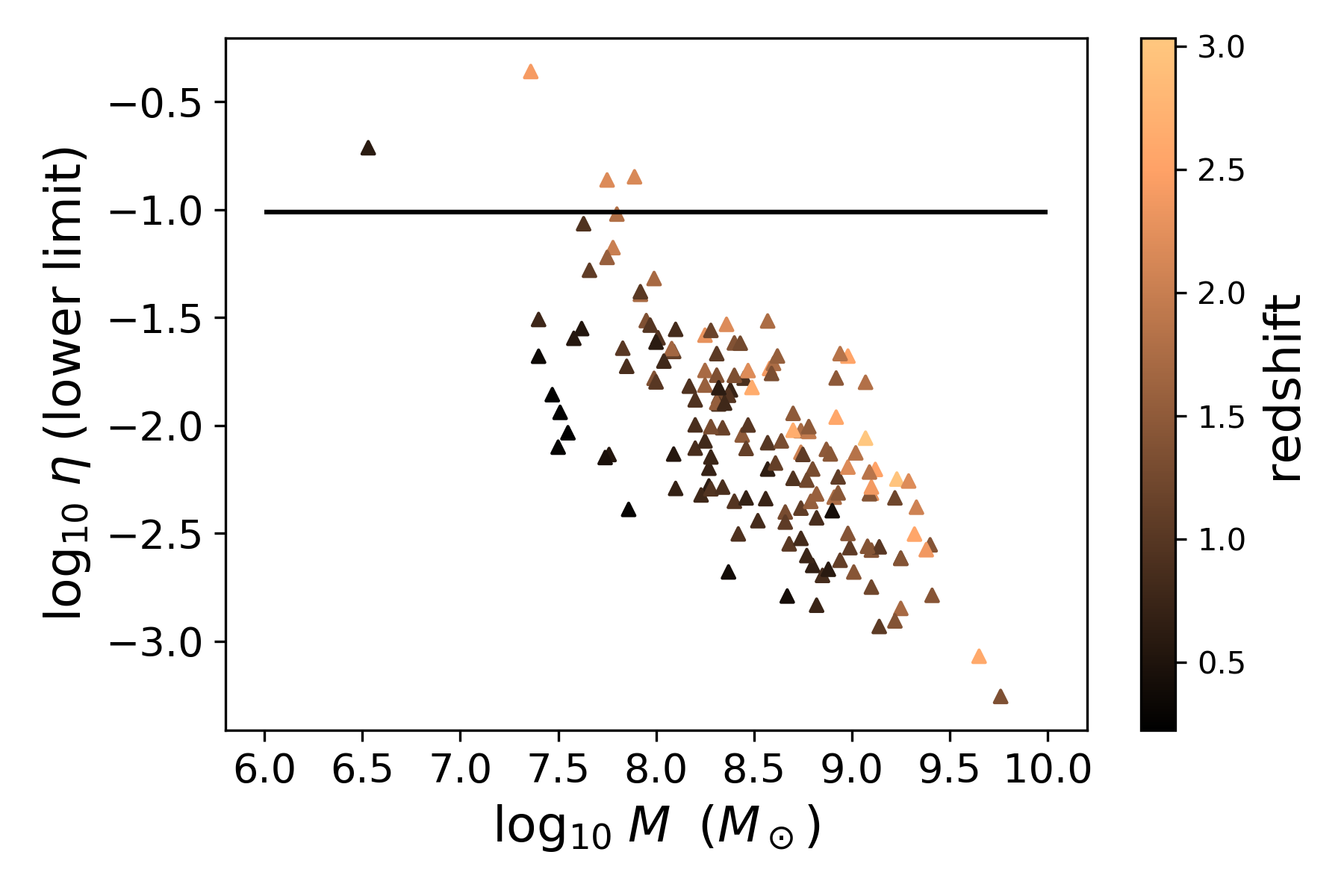}
\caption{Lower limits on the jet efficiencies of the FSRQs. The horizontal line corresponds to the limiting efficiency assuming a maximum allowed spin and a disc thickness $h = 0.13$ in Eq. \ref{eq:eta_avara}.}
\label{fig:eta_mbh_min}
\end{figure}

We show in Fig. \ref{fig:eta_mbh_min} the lower limits on the jet efficiencies of the FSRQs. The horizontal line is the limiting efficiency assuming a black hole rotating with the maximum allowed spin, $a = 0.998$, and also assuming a disc thickness $h = 0.13$. Even considering the lower limits, there are still four objects with efficiencies which seem too high to be explained by the MAD model. We will consider possible explanations for this in the discussion.

\begin{figure}
\includegraphics[width=\linewidth]{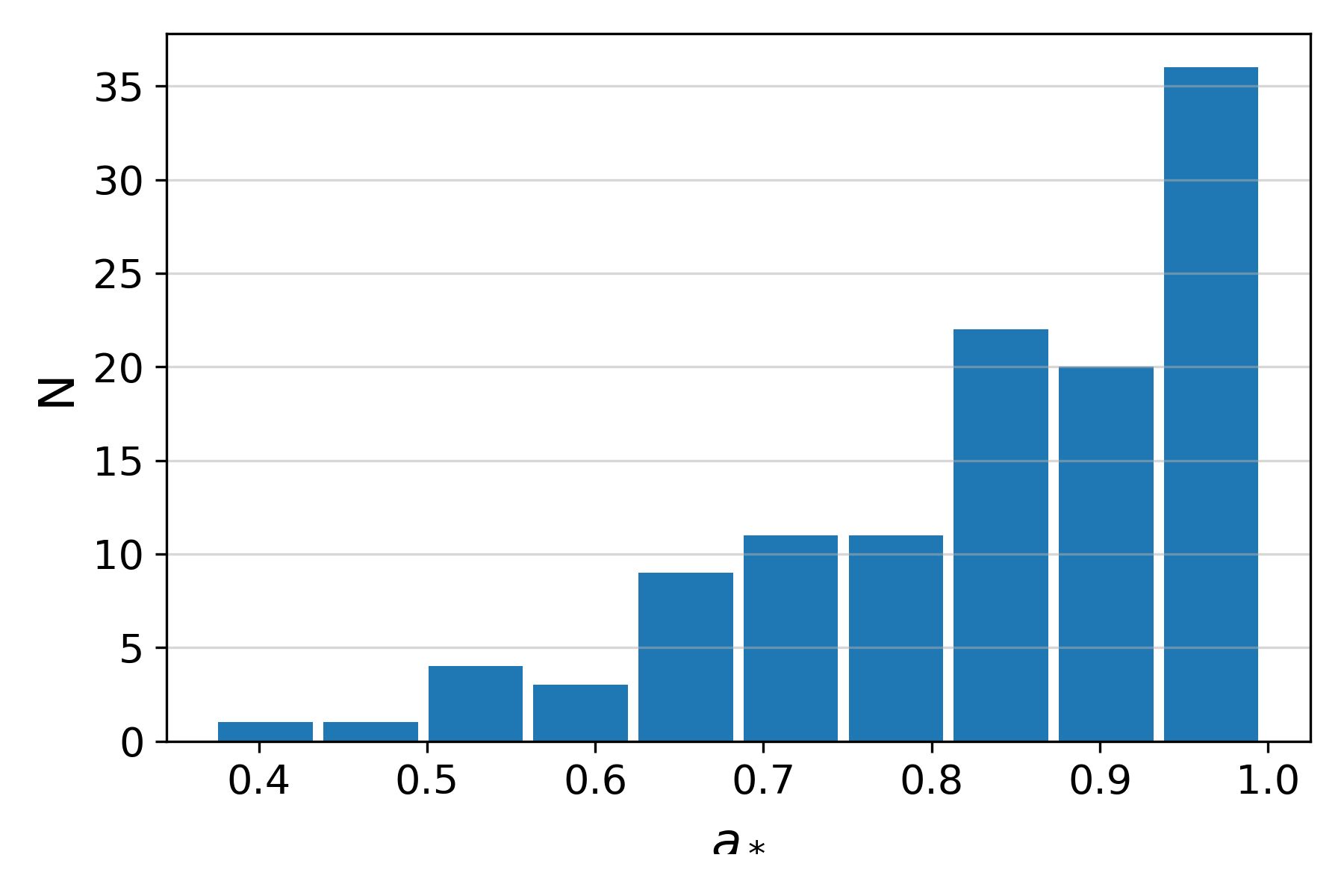}
\caption{Distribution of spins.}
\label{fig:spin_hist}
\end{figure}

\begin{figure}
\includegraphics[width=\linewidth]{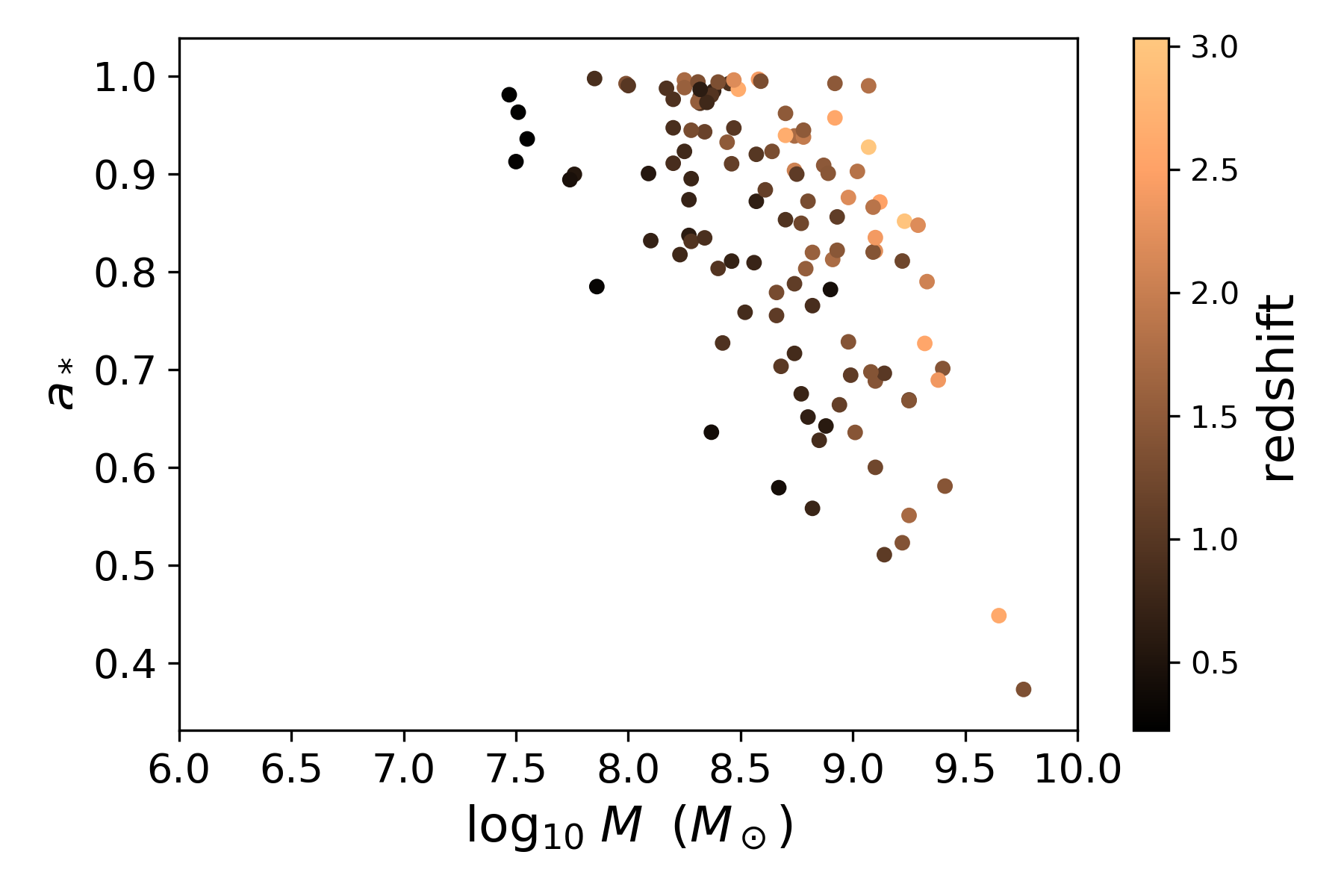}
\caption{Spin estimates for the 118 FSRQs which have efficiencies consistent with the model we adopted.}
\label{fig:spin_mbh}
\end{figure}

For those objects which have $\eta < \rm{max} (\eta_{\rm model})$, we found an average spin of $\langle a_* \rangle = 0.84^{+0.11}_{-0.25}$. The distribution of spins is plotted in Fig. \ref{fig:spin_hist}, while Fig. \ref{fig:spin_mbh} plots the spins as a function of the BH masses. We can see the tendency that the bigger the mass of the SMBH, the lower the associated spin. Because the values of $a_*$ are inferred from $\eta$, the same caveats involved in the $\eta-M$ correlation analysis also apply here: namely that a large degree of the anticorrelation between $a_*$ and $M$ occurs by construction.

\begin{figure}
\includegraphics[width=\linewidth]{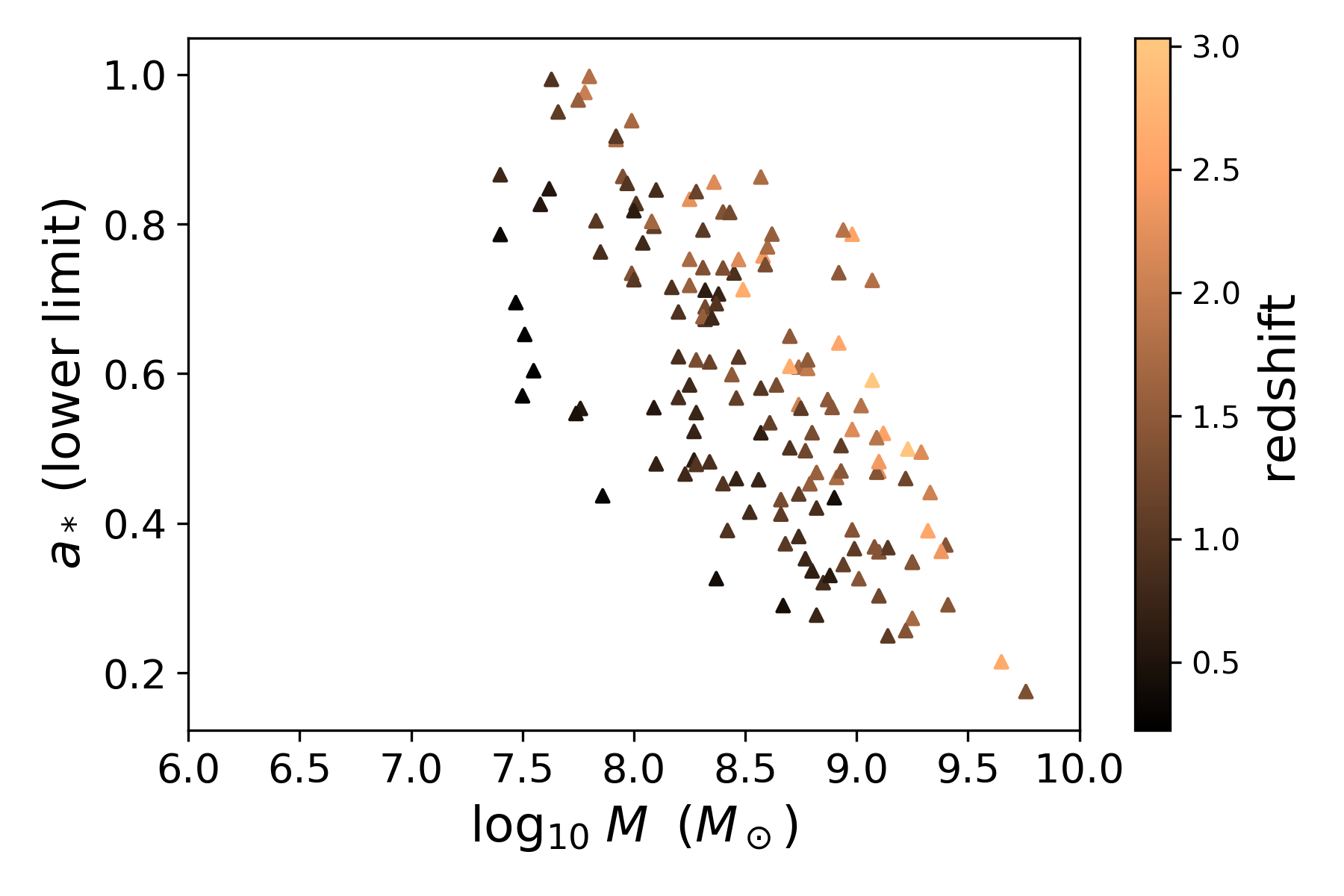}
\caption{Spin lower limits as a function of the mass.}
\label{fig:spin_mbh_min}
\end{figure}

\begin{figure}
\includegraphics[width=\linewidth]{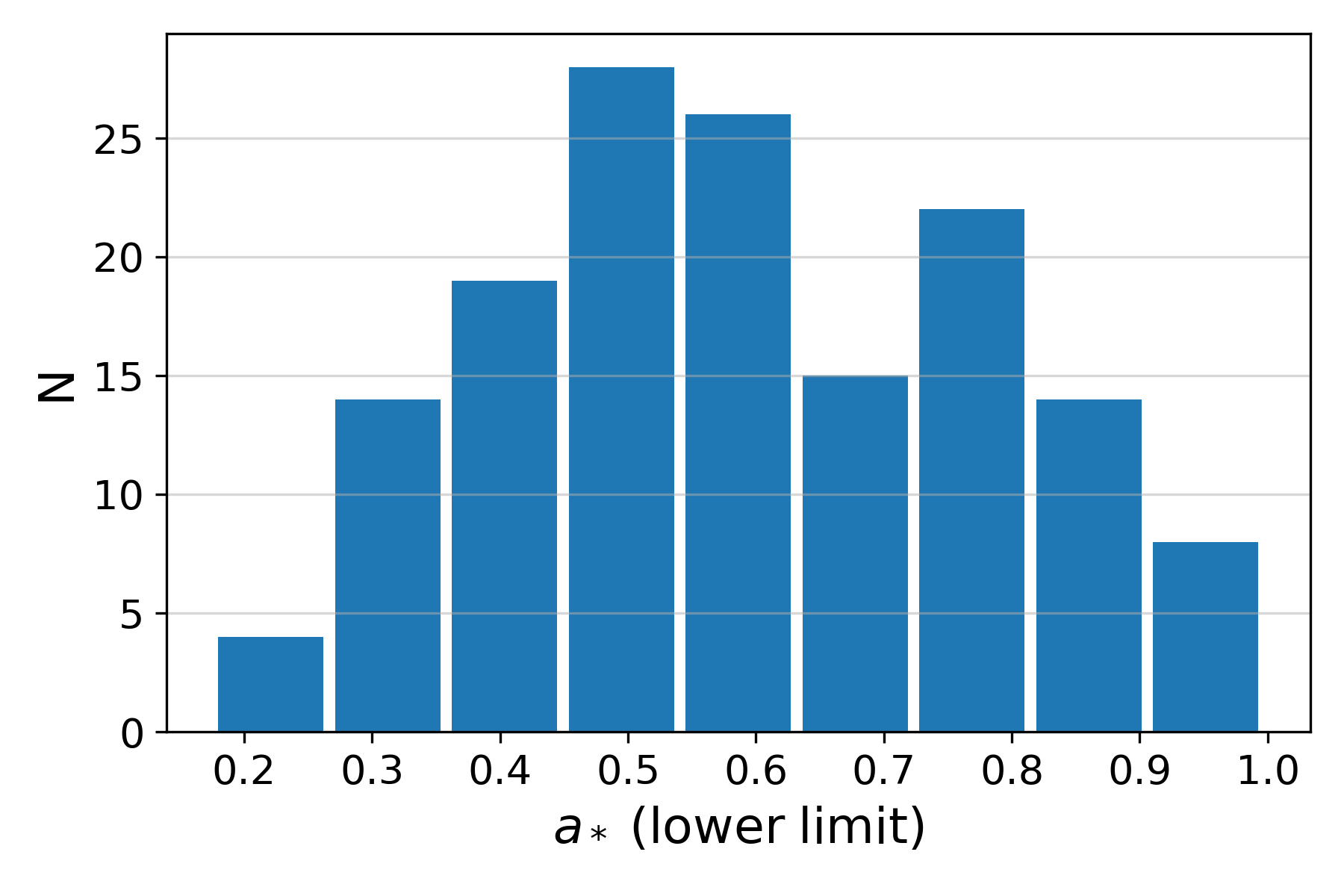}
\caption{Distribution of spin lower limits.}
\label{fig:spin_hist_min}
\end{figure}

The large uncertainties in $\eta$ imply a somewhat large uncertainty of $\approx 0.2$ in the estimated spins. Thus, we decided to concentrate on the lower limits of our spins $a_*$, which average at $0.59$. Figure \ref{fig:spin_mbh_min} displays the relation between the BH masses and the $1\sigma$ lower limit on $a_*$. Figure \ref{fig:spin_hist_min} displays the distribution of lower limits on $a_*$.

We find that even considering the lower limits, 4 out of 154 objects (3 per cent) still have efficiencies which are too high for the model considered. These high-efficiency blazars are listed in Table \ref{tab:higheta}.

\begin{table}
\caption{The four high-efficiency blazars that cannot be explained by the simulation-based model considered.}
\label{tab:higheta}
\begin{tabular}{@{} *{5}c @{}}
\hline
Object name & $z$ & $\log_{10}M$ & $\eta$ & $\eta$\\
(4FGL) & & $M_\odot$ & & (lower limit) \\
\hline
4FGL J0217.0-0821 & 0.607 & 6.53 & 0.986 & 0.194 \\
4FGL J0449.1+1121 & 2.153 & 7.89 & 0.723 & 0.142 \\
4FGL J0601.1-7035 & 2.409 & 7.36 & 2.225 & 0.437 \\
4FGL J2121.0+1901 & 2.18 & 7.75 & 0.699 & 0.137 \\
\hline
\end{tabular}
\end{table}

\section{Discussion}	\label{sec:disc}

There are several methods to estimate the jet power such as SED fitting \citep{Ghisellini2014,Ghisellini09hpb}, radio lobes emission assuming equipartition \citep{Willott99} and radio-core shift \citep{Lobanov98,Shabala12}. \cite{Pjanka17} have shown that these different methods may yield a discrepancy of about one order of magnitude in the final estimates, although they could not conclude which method yields the most accurate result. Methods based on extended observations use average values taken over a larger timescale, whereas SED fitting methods use instantaneous -- and often higher -- values obtained in a short time frame \eg{Pjanka17}.

\begin{figure}
\centering
\includegraphics[width=\linewidth]{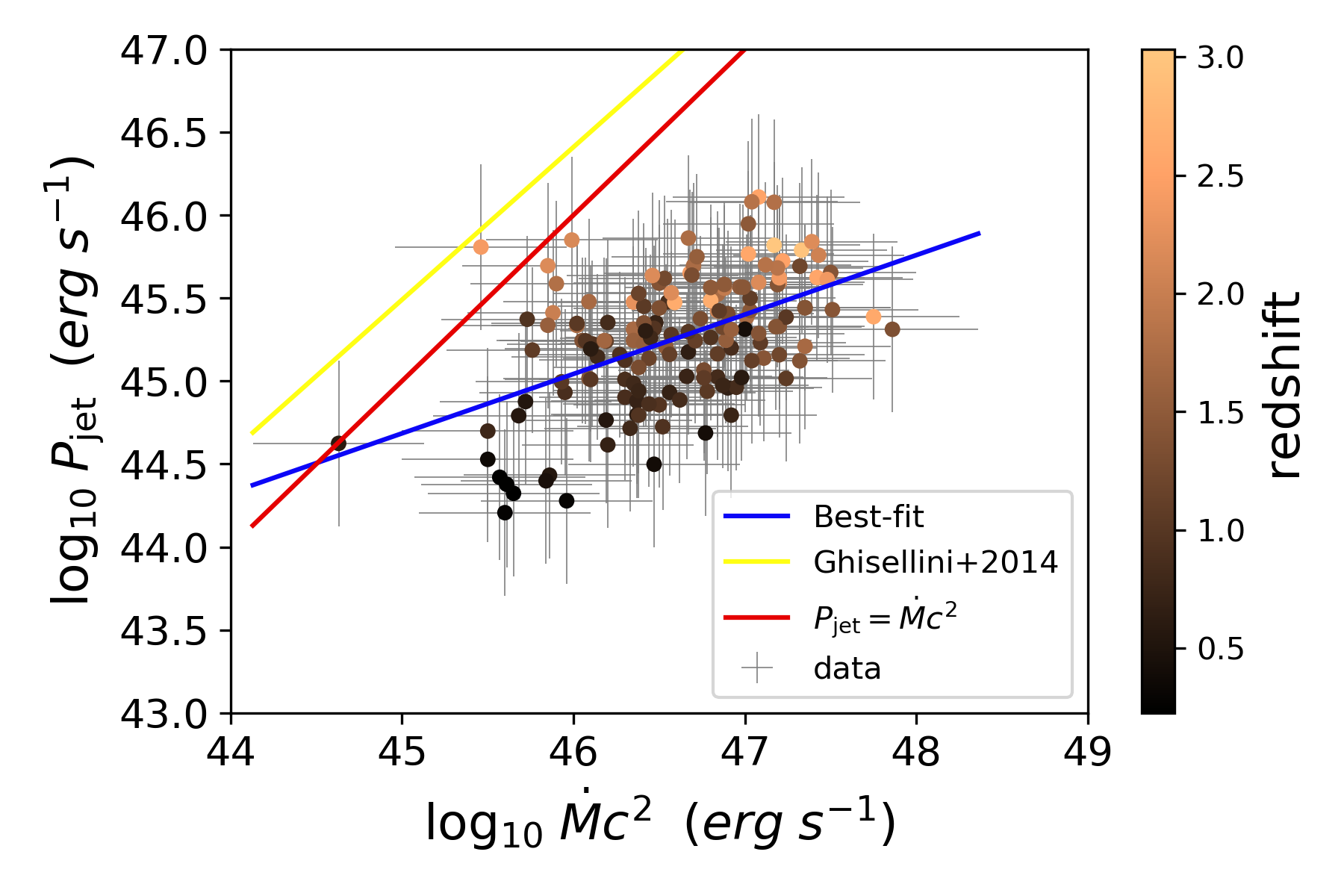}
\caption{Jet powers versus the accretion power $\dot{M}c^2$. The Solid blue line indicates the best-fit relation of G14 and the dashed red line corresponds to $P=\dot{M} c^2$. The points are systematically below the G14 due to the method that we used to constrain $P$.}
\label{fig:pjet_mdot}
\end{figure}

Our method for deriving the jet power is based on the $L_\gamma - P$ correlation of \cite{Nemmen12} which is calibrated based on energetics of giant X-ray cavities inflated by jets over long period of time in radio galaxies \citep{Cavagnolo2010,Meyer2011}. Therefore, our method will systematically give lower values of $P$ reflecting the long-term average jet power, which are often lower than instantaneous estimates. For this reason, our $P$-estimates are lower than those of G14, who reported that the jet power is larger than the accretion power $P_\mathrm{acc} \equiv \dot{M}c^2$ by a factor of $\sim 10$. This is clearly seen in Fig. \ref{fig:pjet_mdot} which compares our results with the best-fit relation of G14. 

In this work, our method for estimating the jet power relies on extended radio luminosities \citep{Nemmen12}, and we found a mean jet efficiency of $0.096$, with a relatively high uncertainty of $0.71\;\mathrm{dex}$. It is interesting to note that, despite methods relying on extended radio luminosities having produced lower jet efficiencies, our results fall in the lower end of blazar studies that used SED fitting methods (i.e. G14), while being on average one order of magnitude higher than the values obtained by \cite{Inoue17}, who also used extended radio luminosities, as in \cite{Willott99}. We note, however, that within one sigma, our results for the jet efficiency are also compatible with those of \cite{Inoue17} and \cite{van-Velzen13}, although our findings of $a_* = 0.84^{+0.11}_{-0.25}$ are higher than the spin distribution found in \cite{Inoue17}, who found $a_* = 0.13^{+0.11}_{-0.059}$.

\cite{Pjanka17} analysed a sample of blazars -- instead of a broader sample of objects classified simply as radio-loud quasars -- and found that jet efficiencies obtained through SED fitting methods are about 10 times larger than those obtained via radio-lobe methods. The presence of $e^{\pm}$ pairs, for example, leads to a decrease in the jet power, and consequently in the jet efficiency. Moreover, radio-lobe methods rely on extended, long-term measurements, and the high variability of blazars over time could be playing a crucial role in leading to smaller values of jet efficiency.

We note that the adoption of larger estimates of $P_\mathrm{jet}$ in this work, approaching those of G14, would result in a systematic increase in the values of $a_*$ in the sample (see Eq. \ref{eq:eta_avara}). Moreover, if the accretion rates are systematically lower than the value we assumed -- $\dot{M} = 0.01\;\dot{M}_\mathrm{Edd}$ instead of $\dot{M} = 0.1\;\dot{M}_\mathrm{Edd}$ -- this would result in an increase in $\eta$, which would, in turn, also cause an increase in the spin values.

Indeed, if we adopt the jet powers as those estimated by G14, we find that the jet efficiencies of \textit{all} objects are larger than the maximum efficiency allowed by the \cite{Avara2016} model. If the jet powers of G14 reflect the true powers of blazars, then the jet model needs to be revised, otherwise the BH would not be able to power these jets. Either the accretion flow is systematically thicker and/or the accretion rates are systematically larger than we considered as discussed above (cf. also section \ref{sec:extremes}).

\subsection{Relation with SMBH and galaxy evolution}{\label{sec:disc_galev}}

Our results support larger values of spins, similar to what has been found in models of SMBH evolution and growth through mergers -- e.g. \cite{Volonteri05,Volonteri07,Volonteri13} --, although not as many as 70 per cent of our black holes can be said to be maximally spinning, or close to it, as suggested by \cite{Volonteri05}. To investigate in more detail possible patterns in the spin evolution of the black holes in our sample, we divided them according to mass ranges and redshifts.

Fig. \ref{fig:spin_mbh} shows the spins distributed according to SMBH mass. We see a general tendency for the spins to decrease as the SMBH masses increase. Such a tendency has been suggested by \cite{King08}, who argued that a series of accretion episodes, in which the discs are randomly oriented due to self-gravity, would spin down the SMBHs to around $0.1$ to $0.3$, and also that mergers would have a short-lived influence on the spin values. However, even if our results point towards a decrease in spin as the mass increases, we note that we have obtained spins which are, on average, higher than those in \cite{King08}. Moreover, while our results suggest moderate to high spins, we also found that some objects considered here display higher values of spins. These could result from mergers with black holes of similar mass and spin orientation. Such mergers are expected to have occurred at some point in the formation of giant elliptical galaxies.

This pattern of spin decreasing as the mass increases has also been found in cosmological simulations by \cite{Dubois14}, although they argue that the accretion disc's angular momentum is conserved, with spin decreasing due to accretion not being significant. Moreover, \cite{Dubois14} also found that mass increases at lower redshifts would be due to more mergers, while accretion would be unable to spin up black holes at lower redshifts due to the gas in the accretion disc being quenched.

We note that \cite{Dubois14} found that spins are high regardless of redshift. In particular, they found that for $z > 2$ the spins are close to the maximum value, while at redshifts between $1$ and $2$, the spins decrease as the black holes undergo more mergers with other black holes of different spin orientations -- and their masses increase -- and accretion can no longer spin up the black holes. In our sample, we found no correlation between spins and redshifts, hence our results do not allow us to reach any substantial conclusions regarding a possible relation between these two quantities.

\subsection{The extremes of blazar jets}    \label{sec:extremes}

Four out of 154 objects in the sample (3 per cent) display jets which are too powerful to be accounted by the MAD thin disc model of \cite{Avara2016} with $h=0.13$ (cf. Table \ref{tab:higheta}). Their efficiencies range between $0.7$ and $2.2$. 

One possibility to conciliate the model with the observations of these extreme blazars is if their accretion rates are higher than the value we assumed. For instance, if they have $\dot{M} \ga 0.7\;\dot{M}_{\rm Edd}$, then their efficiencies are lower than $\eta=0.1$ and can be explained by the jet model with $h=0.13$. Alternatively, we can also explain their efficiencies keeping $\dot{M} = 0.1\;\dot{M}_{\rm Edd}$ and varying $h$ between $0.35$ and $0.67$ (or between $0.16$ and $0.28$ if we consider only the lower limits for $\eta$). These two possibilities might be related since as we increase $\dot{M}$, radiation pressure becomes more important and should make the disc thicker. 

Finally, another possibility if these extreme blazars are accreting at 10 per cent Eddington level and have $h \leq 0.1$ is that the Avara et al. model is inappropriate for them. This is discussed in the subsection that follows.

\subsection{Caveats}{\label{sec:disc_caveats}}

One of the main results of this work is that GRMHD models of jet formation based on moderately thin accretion discs, in which the magnetic field threading the black hole horizon reaches the maximum value sustainable by the accreting matter, are able to explain the energetics of the majority of jetted beamed quasars. 

Standard theory predicts a disc thickness $H/R \sim 0.01$ \eg{Shakura1973, Abramowicz13}. Our assumption of $H/R = 0.13$ is based on the availability of GRMHD simulations of moderately thin discs \citep{Avara2016}. If FSRQs prefer very thin discs with $H/R < 0.1$, then this would imply that eq. \ref{eq:eta_avara} does not describe adequately the energetics of jetted quasars. For instance, if we instead adopt $H/R = 0.01$ we have $\rm{max} (\eta_{\rm model}) = 0.0006$ which would only be able to explain one single object in the sample, even considering the lower limits on $\eta$.

One way of reconciling the observations of powerful jetted quasars with the work of \cite{Avara2016} is if the discs in quasars are indeed only moderately thin, perhaps due to slightly larger accretion rates than usually assumed, which would increase the radiation pressure effects and bloat up the discs similarly to supercritical discs. Higher-resolution simulations of very thin discs such as those of \citep{Liska19} also including radiation pressure effects should help to clarify these issues. 

Finally, we stress that one should be careful when interpreting the anticorrelations between (i) $\eta$ and $M$ (Figure \ref{fig:eta-mass}), (ii) $a_*$ and $M$ (Figure \ref{fig:spin_mbh}) as physical trends. As described in section \ref{sec:res}, the anticorrelation $\eta \propto M^{-1}$ occurs by construction in our work due to our assumptions in estimating the jet efficiencies. Importantly, constraints of $\eta$ which do not depend on $M$ are needed.

\section{Conclusions}   \label{sec:conc}

In this work, we analysed a sample of 154 beamed jetted quasars -- flat spectrum radio quasars -- with gamma-ray luminosities observed with \fermi-LAT and black hole masses estimated with a variety of methods. We used the beamed gamma luminosities as a proxy of their total jet powers and adopted standard quasar accretion rates of 10 per cent Eddington, which  allowed us to estimate their mass accretion rates from the BH masses. Our goal in this work was to test whether current models of jet formation from thin discs are able to explain the energetics in the sample and constrain the BH spin necessary to account for the jet powers. Our main conclusions can be summarised as follows:

(i) A median of 4 per cent of the rest-mass energy associated with accreted matter is converted to jet power in jetted quasars, with the extremes being 0.3 per cent and 200 per cent on the low and high ends, respectively.

(ii)  We find a correlation between BH mass and the observed $\gamma$-ray luminosity. This implies that $L_\gamma$ could potentially be used as a mass estimator in blazar $\gamma$-ray studies: $M = 4 \times 10^8  \left( L_\gamma/10^{47} {\rm erg \ s}^{-1} \right)^{0.37} M_\odot$. The resulting masses have an uncertainty of 0.5 dex.

(iii) Magnetically arrested, moderately thin ($H/R \sim 0.1$) accretion discs around Kerr BHs are able to explain the energetics of the majority of jetted beamed quasars.

(iv) 3 per cent of the sample -- four blazars -- have efficiencies which are too high to be accounted by the model without significant changes to the parameters. Either their accretion rates are higher than assumed, the disc thickness is larger, or the jet model we adopted is inappropriate for them.

(v) By modeling our FSRQ sample with the MAD thin disc model, we obtained an average spin of 0.84. The lowest spin on the sample within $1\sigma$ is 0.2.

(vi) Our spin estimates are compatible with results from models for the cosmological merger-driven evolution of SMBHs which support rapidly rotating black holes.

(vii) If accretion discs in quasars are considerable thinner than assumed here -- i.e. if the discs have $H/R = 0.01$ or lower -- then MAD discs are unable to explain the energetics of the majority of the sample. If this is the case, then general relativistic models for jet production in quasars are missing an important ingredient and need considerable revision.

\section*{Acknowledgements}

We thank Zulema Abraham, Fabio Cafardo, Raniere de Menezes, and Roderik Overzier for helpful discussions. We also thank the anonymous referee for their valuable comments which improved our manuscript. GS is supported by CAPES under grant 88882.332922/2018-01. RN acknowledges support by FAPESP (Funda\c{c}\~ao de Amparo \`a Pesquisa do Estado de S\~ao Paulo) under grant 2017/01461-2.

\bibliographystyle{mnras}
\bibliography{refs,refs-updated} % if your bibtex file is called refs.bib
\appendix

\section{Data}

Table \ref{tab:data} contains the data used in this work.

\begin{table*}
\caption{Data on the objects used in this work. Spins denoted by one asterisk (*) could not be obtained using the jet efficiency $\eta$ as input to the simulation-based model in Equation \ref{eq:eta_avara}, while minimum spins denoted by two asterisks (**) could not be obtained using the same equation for the minimum jet efficiency $\eta_\mathrm{min}$ as input. The uncertainties in mass and jet power are $0.5$ dex.}
\label{tab:data}
\begin{tabular}{@{} *{10}c @{}}
\hline
Object name & Object name & $z$ & $\log_{10}M$ & $\log_{10}L_\gamma$ & $\log_{10}P_\mathrm{jet}$ & $\eta$ & $\eta$ & $a_*$ & $a_*$ \\
(4FGL) & (G14/G15) & & $M_\odot$ & $\mathrm{erg\;s}^{-1}$ & $\mathrm{erg\;s}^{-1}$ & & (lower limit) & & (lower limit) \\
\hline
4FGL J0004.4-4737 & 0004-4736 & 0.88 & 7.85 & 46.53 & 44.93 & 0.096 & 0.019 & 0.997 & 0.763 \\
4FGL J0011.4+0057 & 0011+0057 & 1.493 & 8.44 & 47.07 & 45.2 & 0.046 & 0.009 & 0.932 & 0.599 \\
4FGL J0016.5+1702 & 0015+1700 & 1.709 & 9.25 & 47.08 & 45.21 & 0.007 & 0.001 & 0.551 & 0.273 \\
4FGL J0017.5-0514 & 0017-0512 & 0.226 & 7.55 & 45.34 & 44.32 & 0.047 & 0.009 & 0.936 & 0.604 \\
4FGL J0023.7+4457 & 0023+4456 & 2.023 & 7.78 & 47.47 & 45.41 & 0.34 & 0.067 & * & 0.976 \\
4FGL J0024.7+0349 & 0024+0349 & 0.545 & 7.76 & 45.55 & 44.43 & 0.037 & 0.007 & 0.899 & 0.554 \\
4FGL J0042.2+2319 & 0042+2320 & 1.426 & 9.01 & 46.94 & 45.14 & 0.011 & 0.002 & 0.636 & 0.326 \\
4FGL J0043.8+3425 & 0043+3426 & 0.966 & 8.01 & 47.11 & 45.22 & 0.13 & 0.025 & * & 0.827 \\
4FGL J0044.2-8424 & 0044-8422 & 1.032 & 8.68 & 46.55 & 44.94 & 0.014 & 0.003 & 0.703 & 0.373 \\
4FGL J0047.9+2233 & 0048+2235 & 1.161 & 8.34 & 46.93 & 45.14 & 0.05 & 0.01 & 0.943 & 0.616 \\
4FGL J0050.4-0452 & 0050-0452 & 0.922 & 8.2 & 46.91 & 45.13 & 0.067 & 0.013 & 0.976 & 0.682 \\
4FGL J0058.4+3315 & 0058+3311 & 1.369 & 7.99 & 46.7 & 45.02 & 0.084 & 0.017 & 0.992 & 0.734 \\
4FGL J0102.4+4214 & 0102+4214 & 0.874 & 8.2 & 46.69 & 45.01 & 0.051 & 0.01 & 0.947 & 0.623 \\
4FGL J0102.8+5824 & 0102+5824 & 0.644 & 8.57 & 47.01 & 45.18 & 0.032 & 0.006 & 0.872 & 0.521 \\
4FGL J0104.8-2416 & 0104-2416 & 1.747 & 8.91 & 47.42 & 45.39 & 0.024 & 0.005 & 0.812 & 0.461 \\
4FGL J0157.7-4614 & 0157-4614 & 2.287 & 8.25 & 47.6 & 45.48 & 0.134 & 0.026 & * & 0.833 \\
4FGL J0217.0-0821 & 0217-0820 & 0.607 & 6.53 & 45.93 & 44.62 & 0.986 & 0.194 & * & ** \\
4FGL J0226.5+0938 & 0226+0937 & 2.605 & 9.65 & 47.43 & 45.39 & 0.004 & 0.001 & 0.448 & 0.215 \\
4FGL J0237.8+2848 & 0237+2848 & 1.206 & 9.22 & 48.02 & 45.69 & 0.024 & 0.005 & 0.811 & 0.46 \\
4FGL J0245.4+2408 & 0245+2405 & 2.243 & 9.1 & 47.83 & 45.59 & 0.025 & 0.005 & 0.821 & 0.469 \\
4FGL J0245.9-4650 & 0246-4651 & 1.385 & 8.4 & 47.82 & 45.59 & 0.123 & 0.024 & * & 0.816 \\
4FGL J0252.8-2219 & 0252-2219 & 1.419 & 9.4 & 47.95 & 45.65 & 0.014 & 0.003 & 0.701 & 0.371 \\
4FGL J0253.9+5103 & 0253+5102 & 1.732 & 8.74 & 47.69 & 45.52 & 0.048 & 0.009 & 0.939 & 0.609 \\
4FGL J0257.9-1215 & 0257-1212 & 1.391 & 9.22 & 46.91 & 45.12 & 0.006 & 0.001 & 0.523 & 0.257 \\
4FGL J0303.6-6211 & 0303-6211 & 1.348 & 9.76 & 47.28 & 45.31 & 0.003 & 0.001 & 0.373 & 0.175 \\
4FGL J0309.9-6058 & 0309-6058 & 1.479 & 8.87 & 47.78 & 45.57 & 0.039 & 0.008 & 0.909 & 0.565 \\
4FGL J0315.9-1033 & 0315-1031 & 1.565 & 7.75 & 47.33 & 45.34 & 0.306 & 0.06 & * & 0.966 \\
4FGL J0325.7+2225 & 0325+2224 & 2.066 & 9.33 & 48.15 & 45.76 & 0.021 & 0.004 & 0.79 & 0.441 \\
4FGL J0413.1-5332 & 0413-5332 & 1.024 & 7.83 & 46.66 & 45.0 & 0.116 & 0.023 & * & 0.804 \\
4FGL J0422.1-0644 & 0422-0643 & 0.242 & 7.47 & 45.53 & 44.42 & 0.071 & 0.014 & 0.981 & 0.695 \\
4FGL J0438.4-1254 & 0438-1251 & 1.285 & 8.66 & 46.8 & 45.07 & 0.02 & 0.004 & 0.779 & 0.432 \\
4FGL J0442.6-0017 & 0442-0017 & 0.845 & 8.1 & 47.36 & 45.35 & 0.142 & 0.028 & * & 0.846 \\
4FGL J0449.1+1121 & 0449+1121 & 2.153 & 7.89 & 48.33 & 45.85 & 0.723 & 0.142 & * & ** \\
4FGL J0456.6-3136 & 0456-3136 & 0.865 & 8.2 & 46.48 & 44.9 & 0.04 & 0.008 & 0.911 & 0.568 \\
4FGL J0507.7-6104 & 0507-6104 & 1.089 & 8.74 & 46.99 & 45.16 & 0.021 & 0.004 & 0.788 & 0.439 \\
4FGL J0509.4+1012 & 0509+1011 & 0.621 & 8.27 & 46.27 & 44.8 & 0.027 & 0.005 & 0.837 & 0.485 \\
4FGL J0526.2-4830 & 0526-4830 & 1.3 & 8.8 & 47.46 & 45.41 & 0.032 & 0.006 & 0.872 & 0.521 \\
4FGL J0532.6+0732 & 0532+0732 & 1.254 & 8.43 & 47.88 & 45.62 & 0.123 & 0.024 & * & 0.815 \\
4FGL J0532.9-8325 & 0533-8324 & 0.784 & 7.4 & 46.08 & 44.7 & 0.158 & 0.031 & * & 0.866 \\
4FGL J0533.3+4823 & 0533+4822 & 1.16 & 9.25 & 47.53 & 45.44 & 0.012 & 0.002 & 0.669 & 0.348 \\
4FGL J0541.6-0541 & 0541-0541 & 0.838 & 8.74 & 46.71 & 45.02 & 0.015 & 0.003 & 0.716 & 0.382 \\
4FGL J0601.1-7035 & 0601-7036 & 2.409 & 7.36 & 48.25 & 45.81 & 2.225 & 0.437 & * & ** \\
4FGL J0608.0-0835 & 0607-0834 & 0.87 & 8.82 & 47.06 & 45.2 & 0.019 & 0.004 & 0.765 & 0.42 \\
4FGL J0608.1-1521 & 0608-1520 & 1.094 & 8.09 & 47.14 & 45.24 & 0.112 & 0.022 & * & 0.797 \\
4FGL J0625.8-5441 & 0625-5438 & 2.051 & 8.74 & 47.5 & 45.42 & 0.038 & 0.008 & 0.904 & 0.559 \\
4FGL J0654.4+4514 & 0654+4514 & 0.928 & 8.17 & 46.98 & 45.16 & 0.077 & 0.015 & 0.987 & 0.715 \\
4FGL J0654.3+5042 & 0654+5042 & 1.253 & 8.32 & 47.17 & 45.26 & 0.069 & 0.014 & 0.979 & 0.689 \\
4FGL J0713.8+1935 & 0713+1935 & 0.54 & 7.62 & 46.42 & 44.88 & 0.143 & 0.028 & * & 0.847 \\
4FGL J0721.3+0405 & 0721+0406 & 0.665 & 8.8 & 46.59 & 44.96 & 0.011 & 0.002 & 0.651 & 0.336 \\
4FGL J0723.5+2900 & 0723+2859 & 0.966 & 8.4 & 46.38 & 44.86 & 0.023 & 0.004 & 0.803 & 0.453 \\
4FGL J0725.2+1425 & 0725+1425 & 1.038 & 8.31 & 47.55 & 45.45 & 0.11 & 0.022 & * & 0.792 \\
4FGL J0746.4+2546 & 0746+2549 & 2.979 & 9.23 & 48.21 & 45.79 & 0.029 & 0.006 & 0.852 & 0.499 \\
4FGL J0805.4+6147 & 0805+6144 & 3.033 & 9.07 & 48.27 & 45.82 & 0.044 & 0.009 & 0.927 & 0.591 \\
4FGL J0824.7+5552 & 0825+5555 & 1.418 & 9.1 & 47.31 & 45.33 & 0.013 & 0.003 & 0.688 & 0.362 \\
4FGL J0830.8+2410 & 0830+2410 & 0.942 & 8.7 & 47.18 & 45.26 & 0.029 & 0.006 & 0.853 & 0.501 \\
4FGL J0909.1+0121 & 0909+0121 & 1.026 & 9.14 & 47.42 & 45.38 & 0.014 & 0.003 & 0.696 & 0.367 \\
4FGL J0910.6+2247 & 0910+2248 & 2.661 & 8.7 & 47.62 & 45.49 & 0.048 & 0.01 & 0.939 & 0.61 \\
4FGL J0912.2+4127 & 0912+4126 & 2.563 & 9.32 & 47.89 & 45.62 & 0.016 & 0.003 & 0.727 & 0.39 \\
4FGL J0920.9+4441 & 0920+4441 & 2.189 & 9.29 & 48.31 & 45.84 & 0.028 & 0.006 & 0.847 & 0.495 \\
4FGL J0921.6+6216 & 0921+6215 & 1.453 & 8.93 & 47.5 & 45.42 & 0.025 & 0.005 & 0.822 & 0.47 \\
\hline
\end{tabular}
\end{table*}

\begin{table*}
\contcaption{}
\label{tab:datacont}
\begin{tabular}{@{} *{10}c @{}}
\hline
Object name & Object name & $z$ & $\log_{10}M$ & $\log_{10}L_\gamma$ & $\log_{10}P_\mathrm{jet}$ & $\eta$ & $\eta$ & $a_*$ & $a_*$ \\
(4FGL) & (G14/G15) & & $M_\odot$ & $\mathrm{erg\;s}^{-1}$ & $\mathrm{erg\;s}^{-1}$ & & (lower limit) & & (lower limit) \\
\hline
4FGL J0924.0+2816 & 0923+2815 & 0.744 & 8.82 & 46.26 & 44.79 & 0.007 & 0.001 & 0.558 & 0.277 \\
4FGL J0923.5+4125 & 0923+4125 & 1.732 & 7.92 & 47.32 & 45.34 & 0.206 & 0.041 & * & 0.913 \\
4FGL J0937.1+5008 & 0937+5008 & 0.276 & 7.5 & 45.11 & 44.21 & 0.04 & 0.008 & 0.912 & 0.57 \\
4FGL J0946.6+1016 & 0946+1017 & 1.006 & 8.47 & 47.22 & 45.28 & 0.051 & 0.01 & 0.947 & 0.622 \\
4FGL J0949.2+1749 & 0949+1752 & 0.693 & 8.1 & 45.91 & 44.62 & 0.026 & 0.005 & 0.832 & 0.479 \\
4FGL J0956.7+2516 & 0956+2515 & 0.708 & 8.46 & 46.53 & 44.93 & 0.024 & 0.005 & 0.811 & 0.46 \\
4FGL J0957.6+5523 & 0957+5522 & 0.899 & 8.45 & 47.6 & 45.48 & 0.084 & 0.017 & 0.992 & 0.735 \\
4FGL J1012.7+2439 & 1012+2439 & 1.8 & 7.8 & 47.82 & 45.59 & 0.486 & 0.095 & * & 0.997 \\
4FGL J1016.0+0512 & 1016+0513 & 1.714 & 7.99 & 47.6 & 45.48 & 0.244 & 0.048 & * & 0.938 \\
4FGL J1018.4+3540 & 1018+3542 & 1.228 & 9.1 & 46.98 & 45.16 & 0.009 & 0.002 & 0.6 & 0.303 \\
4FGL J1033.9+6050 & 1032+6051 & 1.064 & 8.75 & 47.5 & 45.42 & 0.037 & 0.007 & 0.9 & 0.554 \\
4FGL J1033.1+4115 & 1033+4116 & 1.117 & 8.61 & 47.14 & 45.24 & 0.034 & 0.007 & 0.884 & 0.534 \\
4FGL J1033.9+6050 & 1033+6051 & 1.401 & 9.09 & 47.81 & 45.58 & 0.025 & 0.005 & 0.82 & 0.468 \\
4FGL J1037.7-2822 & 1037-2823 & 1.066 & 8.99 & 47.12 & 45.23 & 0.014 & 0.003 & 0.694 & 0.366 \\
4FGL J1043.2+2408 & 1043+2408 & 0.559 & 8.09 & 46.21 & 44.76 & 0.038 & 0.007 & 0.9 & 0.555 \\
4FGL J1106.0+2813 & 1106+2812 & 0.843 & 8.85 & 46.59 & 44.96 & 0.01 & 0.002 & 0.627 & 0.321 \\
4FGL J1112.5+3448 & 1112+3446 & 1.956 & 8.78 & 47.76 & 45.56 & 0.048 & 0.009 & 0.938 & 0.607 \\
4FGL J1124.0+2336 & 1124+2336 & 1.549 & 8.79 & 47.15 & 45.25 & 0.023 & 0.004 & 0.803 & 0.453 \\
4FGL J1146.9+3958 & 1146+3958 & 1.088 & 8.93 & 47.64 & 45.5 & 0.029 & 0.006 & 0.856 & 0.504 \\
4FGL J1152.3-0839 & 1152-0841 & 2.367 & 9.38 & 47.87 & 45.61 & 0.014 & 0.003 & 0.689 & 0.363 \\
4FGL J1154.0+6018 & 1154+6022 & 1.12 & 8.94 & 46.91 & 45.12 & 0.012 & 0.002 & 0.664 & 0.345 \\
4FGL J1159.5+2914 & 1159+2914 & 0.725 & 8.38 & 47.36 & 45.35 & 0.074 & 0.015 & 0.985 & 0.706 \\
4FGL J1208.9+5441 & 1208+5441 & 1.344 & 8.4 & 47.53 & 45.44 & 0.087 & 0.017 & 0.994 & 0.741 \\
4FGL J1209.8+1810 & 1209+1810 & 0.845 & 8.52 & 46.44 & 44.89 & 0.018 & 0.004 & 0.758 & 0.415 \\
4FGL J1222.5+0414 & 1222+0413 & 0.966 & 8.37 & 47.29 & 45.32 & 0.07 & 0.014 & 0.98 & 0.693 \\
4FGL J1224.9+2122 & 1224+2122 & 0.434 & 8.9 & 47.28 & 45.31 & 0.021 & 0.004 & 0.782 & 0.434 \\
4FGL J1223.9+5000 & 1224+5001 & 1.065 & 8.66 & 46.71 & 45.02 & 0.018 & 0.004 & 0.755 & 0.412 \\
4FGL J1228.7+4858 & 1228+4858 & 1.722 & 8.25 & 47.28 & 45.31 & 0.092 & 0.018 & 0.996 & 0.753 \\
4FGL J1239.5+0443 & 1239+0443 & 1.761 & 8.57 & 48.36 & 45.86 & 0.155 & 0.03 & * & 0.863 \\
4FGL J1257.8+3228 & 1257+3229 & 0.806 & 8.25 & 46.64 & 44.99 & 0.043 & 0.008 & 0.923 & 0.585 \\
4FGL J1303.6-4622 & 1303-4621 & 1.664 & 8.08 & 47.14 & 45.24 & 0.116 & 0.023 & * & 0.803 \\
4FGL J1310.5+3221 & 1310+3220 & 0.997 & 8.57 & 47.25 & 45.3 & 0.042 & 0.008 & 0.92 & 0.581 \\
4FGL J1317.6+3428 & 1317+3425 & 1.055 & 9.14 & 46.7 & 45.02 & 0.006 & 0.001 & 0.51 & 0.25 \\
4FGL J1321.1+2216 & 1321+2216 & 0.943 & 8.32 & 47.11 & 45.23 & 0.064 & 0.013 & 0.972 & 0.673 \\
4FGL J1326.9+2210 & 1327+2210 & 1.403 & 9.25 & 47.53 & 45.44 & 0.012 & 0.002 & 0.668 & 0.348 \\
4FGL J1332.6-1256 & 1332-1256 & 1.492 & 8.78 & 47.81 & 45.58 & 0.05 & 0.01 & 0.945 & 0.618 \\
4FGL J1333.7+5056 & 1333+5057 & 1.362 & 7.95 & 47.14 & 45.24 & 0.156 & 0.031 & * & 0.864 \\
4FGL J1343.6+5755 & 1343+5754 & 0.933 & 8.42 & 46.13 & 44.72 & 0.016 & 0.003 & 0.727 & 0.39 \\
4FGL J1344.2-1723 & 1344-1723 & 2.506 & 9.12 & 48.09 & 45.72 & 0.032 & 0.006 & 0.871 & 0.52 \\
4FGL J1345.5+4453 & 1345+4452 & 2.534 & 8.98 & 48.84 & 46.11 & 0.107 & 0.021 & * & 0.786 \\
4FGL J1347.6-3751 & 1347-3750 & 1.3 & 8.28 & 46.83 & 45.08 & 0.05 & 0.01 & 0.945 & 0.618 \\
4FGL J1350.8+3033 & 1350+3034 & 0.712 & 8.27 & 46.43 & 44.88 & 0.032 & 0.006 & 0.873 & 0.523 \\
4FGL J1358.1+7642 & 1357+7643 & 1.585 & 8.25 & 47.15 & 45.24 & 0.078 & 0.015 & 0.988 & 0.718 \\
4FGL J1359.1+5544 & 1359+5544 & 1.014 & 8.0 & 46.68 & 45.01 & 0.081 & 0.016 & 0.99 & 0.726 \\
4FGL J1423.5-7829 & 1423-7829 & 0.788 & 8.23 & 46.11 & 44.72 & 0.024 & 0.005 & 0.817 & 0.466 \\
4FGL J1436.9+2321 & 1436+2321 & 1.548 & 8.31 & 47.11 & 45.22 & 0.065 & 0.013 & 0.974 & 0.676 \\
4FGL J1438.9+3710 & 1438+3710 & 2.399 & 8.58 & 47.94 & 45.65 & 0.094 & 0.018 & 0.997 & 0.757 \\
4FGL J1441.6-1522 & 1441-1523 & 2.642 & 8.49 & 47.6 & 45.47 & 0.076 & 0.015 & 0.986 & 0.712 \\
4FGL J1443.9+2501 & 1443+2501 & 0.939 & 7.63 & 47.4 & 45.37 & 0.439 & 0.086 & * & 0.993 \\
4FGL J1504.4+1029 & 1504+1029 & 1.839 & 8.94 & 48.79 & 46.08 & 0.11 & 0.022 & * & 0.792 \\
4FGL J1522.1+3144 & 1522+3144 & 1.484 & 8.92 & 48.52 & 45.95 & 0.085 & 0.017 & 0.992 & 0.735 \\
4FGL J1539.6+2743 & 1539+2744 & 2.191 & 8.47 & 47.71 & 45.53 & 0.092 & 0.018 & 0.996 & 0.753 \\
4FGL J1549.5+0236 & 1549+0237 & 0.414 & 8.67 & 46.05 & 44.69 & 0.008 & 0.002 & 0.579 & 0.29 \\
4FGL J1550.7+0528 & 1550+0527 & 1.417 & 8.98 & 47.23 & 45.29 & 0.016 & 0.003 & 0.728 & 0.391 \\
4FGL J1553.6+1257 & 1553+1256 & 1.308 & 8.64 & 47.4 & 45.38 & 0.043 & 0.008 & 0.923 & 0.585 \\
4FGL J1608.7+1029 & 1608+1029 & 1.232 & 8.77 & 47.3 & 45.32 & 0.028 & 0.006 & 0.849 & 0.497 \\
4FGL J1613.6+3411 & 1613+3412 & 1.4 & 9.08 & 47.31 & 45.33 & 0.014 & 0.003 & 0.697 & 0.369 \\
4FGL J1616.6+4630 & 1616+4632 & 0.95 & 8.28 & 46.26 & 44.79 & 0.026 & 0.005 & 0.831 & 0.479 \\
4FGL J1617.3-5849 & 1617-5848 & 1.422 & 9.41 & 47.51 & 45.43 & 0.008 & 0.002 & 0.581 & 0.291 \\
4FGL J1628.8-6149 & 1628-6152 & 2.578 & 8.92 & 48.17 & 45.77 & 0.056 & 0.011 & 0.957 & 0.641 \\
4FGL J1635.2+3808 & 1635+3808 & 1.813 & 9.07 & 48.78 & 46.08 & 0.081 & 0.016 & 0.99 & 0.725 \\
4FGL J1637.7+4717 & 1637+4717 & 0.735 & 8.56 & 46.72 & 45.03 & 0.023 & 0.005 & 0.809 & 0.458 \\
\hline
\end{tabular}
\end{table*}

\begin{table*}
\contcaption{}
\label{tab:datacont2}
\begin{tabular}{@{} *{10}c @{}}
\hline
Object name & Object name & $z$ & $\log_{10}M$ & $\log_{10}L_\gamma$ & $\log_{10}P_\mathrm{jet}$ & $\eta$ & $\eta$ & $a_*$ & $a_*$ \\
(4FGL) & (G14/G15) & & $M_\odot$ & $\mathrm{erg\;s}^{-1}$ & $\mathrm{erg\;s}^{-1}$ & & (lower limit) & & (lower limit) \\
\hline
4FGL J1642.9+3948 & 1642+3940 & 0.593 & 8.88 & 46.71 & 45.02 & 0.011 & 0.002 & 0.642 & 0.33 \\
4FGL J1703.6-6213 & 1703-6212 & 1.747 & 8.6 & 48.03 & 45.7 & 0.099 & 0.019 & 0.998 & 0.769 \\
4FGL J1709.7+4318 & 1709+4318 & 1.027 & 7.92 & 47.35 & 45.35 & 0.212 & 0.042 & * & 0.917 \\
4FGL J1734.3+3858 & 1734+3857 & 0.975 & 7.97 & 47.14 & 45.24 & 0.149 & 0.029 & * & 0.854 \\
4FGL J1736.6+0628 & 1736+0631 & 2.387 & 9.1 & 47.89 & 45.62 & 0.026 & 0.005 & 0.835 & 0.482 \\
4FGL J1802.6-3940 & 1802-3940 & 1.319 & 8.59 & 47.92 & 45.64 & 0.089 & 0.017 & 0.995 & 0.746 \\
4FGL J1818.6+0903 & 1818+0903 & 0.354 & 7.4 & 45.74 & 44.53 & 0.107 & 0.021 & * & 0.786 \\
4FGL J1830.1+0617 & 1830+0619 & 0.745 & 8.77 & 46.62 & 44.97 & 0.013 & 0.002 & 0.675 & 0.353 \\
4FGL J1848.4+3217 & 1848+3219 & 0.8 & 8.04 & 46.95 & 45.15 & 0.101 & 0.02 & 0.999 & 0.775 \\
4FGL J1902.9-6748 & 1903-6749 & 0.254 & 7.51 & 45.45 & 44.38 & 0.059 & 0.012 & 0.963 & 0.652 \\
4FGL J1955.2+1358 & 1955+1358 & 0.743 & 8.28 & 46.55 & 44.94 & 0.036 & 0.007 & 0.895 & 0.548 \\
4FGL J1959.1-4247 & 1959-4246 & 2.178 & 8.98 & 47.83 & 45.59 & 0.033 & 0.006 & 0.876 & 0.525 \\
4FGL J2026.0-2845 & 2025-2845 & 0.884 & 8.34 & 46.4 & 44.86 & 0.026 & 0.005 & 0.834 & 0.482 \\
4FGL J2035.4+1056 & 2035+1056 & 0.601 & 8.0 & 47.05 & 45.19 & 0.124 & 0.024 & * & 0.818 \\
4FGL J2110.3+0808 & 2110+0809 & 1.58 & 8.82 & 47.27 & 45.31 & 0.025 & 0.005 & 0.82 & 0.468 \\
4FGL J2118.0+0019 & 2118+0013 & 0.463 & 7.74 & 45.49 & 44.4 & 0.036 & 0.007 & 0.894 & 0.547 \\
4FGL J2121.0+1901 & 2121+1901 & 2.18 & 7.75 & 48.03 & 45.7 & 0.699 & 0.137 & * & ** \\
4FGL J2135.3-5006 & 2135-5006 & 2.181 & 8.36 & 47.91 & 45.64 & 0.15 & 0.029 & * & 0.856 \\
4FGL J2145.0-3356 & 2145-3357 & 1.361 & 8.31 & 47.35 & 45.35 & 0.087 & 0.017 & 0.994 & 0.742 \\
4FGL J2157.5+3127 & 2157+3127 & 1.448 & 8.89 & 47.78 & 45.57 & 0.038 & 0.007 & 0.901 & 0.555 \\
4FGL J2201.5-8339 & 2202-8338 & 1.865 & 9.09 & 48.0 & 45.68 & 0.031 & 0.006 & 0.866 & 0.514 \\
4FGL J2212.0+2356 & 2212+2355 & 1.125 & 8.46 & 46.98 & 45.16 & 0.04 & 0.008 & 0.91 & 0.567 \\
4FGL J2219.2+1806 & 2219+1806 & 1.071 & 7.66 & 47.03 & 45.19 & 0.267 & 0.052 & * & 0.95 \\
4FGL J2229.7-0832 & 2229-0832 & 1.56 & 8.62 & 48.14 & 45.75 & 0.107 & 0.021 & * & 0.787 \\
4FGL J2236.3+2828 & 2236+2828 & 0.79 & 8.35 & 47.18 & 45.26 & 0.065 & 0.013 & 0.973 & 0.675 \\
4FGL J2237.0-3921 & 2237-3921 & 0.297 & 7.86 & 45.25 & 44.28 & 0.021 & 0.004 & 0.785 & 0.437 \\
4FGL J2244.2+4057 & 2244+4057 & 1.171 & 8.28 & 47.7 & 45.53 & 0.141 & 0.028 & * & 0.843 \\
4FGL J2321.9+3204 & 2321+3204 & 1.489 & 8.7 & 47.77 & 45.56 & 0.058 & 0.011 & 0.962 & 0.65 \\
4FGL J2327.5+0939 & 2327+0940 & 1.841 & 9.02 & 48.04 & 45.7 & 0.038 & 0.007 & 0.902 & 0.557 \\
4FGL J2331.0-2147 & 2331-2148 & 0.563 & 7.58 & 46.26 & 44.79 & 0.129 & 0.025 & * & 0.826 \\
4FGL J2334.2+0736 & 2334+0736 & 0.401 & 8.37 & 45.68 & 44.5 & 0.011 & 0.002 & 0.636 & 0.326 \\
4FGL J2345.2-1555 & 2345-1555 & 0.621 & 8.32 & 47.26 & 45.3 & 0.076 & 0.015 & 0.986 & 0.712 \\
\hline
\end{tabular}
\end{table*}

\bsp	% typesetting comment
\label{lastpage}
\end{document}